\documentclass{aastex631}

\shorttitle{AASTeX v6.31 RS Oph NuSTAR and shocks}
\shortauthors{Orio et al.}
\graphicspath{{./}{figures/}}

\begin{document}

\title{The powerful shocks in RS Oph: NuSTAR X-ray data and a complete review}

\correspondingauthor{Marina Orio}
\email{orio@astro.wisc.edu}

\author[0000-0003-1563-9803]{Marina Orio}
\affiliation{Department of Astronomy, University of Wisconsin 
475 N. Charter Str., Madison, WI, USA}
\affiliation{INAF-Padova, vicolo Osservatorio 5,
35122 Padova, Italy.}
\author[0000-0002-2647-4373]{G. J. M. Luna}
\affil{Universidad Nacional de Hurlingham (UNAHUR). Laboratorio de Investigación y Desarrollo Experimental en Computación, Av. Gdor. Vergara 2222, Villa Tesei, Buenos Aires, Argentina}
\affiliation{Consejo Nacional de Investigaciones Científicas y Técnicas (CONICET).}
\author{Ehud Behar}
\author{Rebecca Diesing}
\author{Jay Gallagher}
\author{Joanna Mikolajewska}
\author{Jan-Uwe Ness}


\begin{abstract}
In the 2021 outburst of RS Ophiuchi, the gamma-ray
 and the X-ray flux were measured quasi-simultaneously from day 1
 after the  optical peak, offering the first comprehensive view of shocks 
in a nova occurring in  a symbiotic system. We present a previously
 unpublished observation done with {\sl NuSTAR} in the
 3-79  keV range, 9 days after maximum, and
 we review the complex history of the
evidence of shocks in the previous outbursts of this nova  in
 the light of the intensive X-ray monitoring of 2021.  
   We find evidence that the shock causing the particle acceleration
 measured with the Cherenkov telescopes produced also the thermal flux 
detected in the 0.2-30 keV
 X-ray range, while the  large gamma-ray flux  observed with  {\sl Fermi}
   after about a day, is not consistent  with the X-ray observations.
 We conclude that an initial, strong shock, with  particle-particle loss
 timescale shorter  than the timescale of particle acceleration at 
energy higher than a few GeV, occurred close to the red giant atmosphere, 
where either the X-rays' emitting volume was  reduced by turbulence, 
or - perhaps less likely - the X-rays were completely absorbed by 
large column density near the giant and by the accretion wake 
along the line of sight. We compare RS Oph with other novae
 in long period systems with evolved companions,  
discussing how the shocks' phenomenology is a powerful tool to derive 
other physical parameters. Finally, we discuss predictions that 
in T CrB, expected to have a new  nova outburst 
within the next few years, the  shocks may not be as energetic as in RS Oph.
\end{abstract}


\section{Introduction \label{sec:intro}}
 {\it Novae} are luminous at all wavelengths from
 gamma-rays to radio \citep[see review by ][]{Chomiuk2021},
 and X-rays have proven to be a very important
 window to understanding their physics since the '80ies \citep{Ogelman1984}.
  Before we continue,
 we clarify that in this article, the definition of {\it ``nova''}
 is used exclusively for large amplitude outbursts of at
 least 8 magnitudes in optical, due to a thermonuclear runaway
 followed by mass ejection in a wind (we are not referring to 
 any phenomena without mass ejection, such as 
 Z And activity \citep[e.g.][]{Sokoloski2006} or millinovae
 \citep{Mroz2024}). 
 An important success of the nova models was the prediction that,
 after a thermonuclear flash, 
 the  WD atmosphere contracts while there is still a burning
 layer near the surface, returning
 almost to the pre-outburst radius \citep{Yaron2005, Starrfield2012, Wolf2013}.
 This is in fact observed: the peak wavelength of the emission,
  after an initial, brief (hours-long) fireball phase in soft X-rays
 (observed unambiguously only once \citep{Konig2022}) decreases
 in the next hours as the photosphere expands to a sort
 of blue-giant configuration, then it shifts again from 
 the optical range to the UV, a little later to the extreme UV, and finally,
 to the soft X-rays, in a phase of constant
 bolometric luminosity powered by shell burning
\citep[see, among others,][]{Balman1998, Orio2012, Balman2025}. The
 central source in most cases is observable as a luminous
 supersoft X-ray source (SSS) after a time ranging from days to few months
 (specifically, about a month for RS Oph),
 with near-Eddington luminosity close to 10$^{38}$ erg s$^{-1}$
 and peak temperature up to a million K  \citep[see][]{Balman2025}.  

 A phenomenon that was observed and could not be included
 in the evolutionary models, however, were the hard
 X-rays observed within days of the outburst.
 The X-ray flux in most cases has an initial peak at energy
 at or above a few keV,  {\it due to shocks in the ejecta} 
\citep[e.g.][]{Balman1998, Nelson2008, Orio2012, Chomiuk2021}.
  In classical novae with a main sequence
 or slightly evolved secondary, the shocks - causing
 X-ray luminosity peaking around 10$^{34}$ erg s$^{-1}$ -
 have been attributed 
 to colliding winds in different phases of the outflow,
 probably a dense and slow equatorial one and a fast polar ``jet'',  
 while in novae that host a red giant and have wide orbital
 separations (at least 1 AU), the maximum X-ray luminosity exceeds
 10$^{36}$ erg s$^{-1}$ and is thought to be due to the ejecta
 shocking the red giant wind, which has much lower velocity (a few thousand
 km s$^{-1}$ vs. a few tens km s$^{-1}$).

RS Oph is arguably the best known recurrent 
 nova in  a symbiotic system.  Symbiotic binaries host a compact object (usually a white dwarf, WD), and a red giant or AGB secondary. Although most nova outbursts are observed in 
short-orbital-period systems (cataclysmic
 variables,  with orbital
 periods of a few hours up to $\simeq$ 2 days), novae also occur in symbiotic
systems, which have years-long orbital periods. The root cause of a nova outburst
 is a thermonuclear runaway (TNR) on the surface of the
 WD, which is accreting material from its binary companion. The TNR is 
 followed by a radiation driven wind, that depletes the
 accreted envelope \citep{Starrfield2012,
 Wolf2013}. In  short-orbital-period novae, but not in the 
long-orbital-period ones, 
 a different, initial mechanism of mass loss 
 is expected to occur before the onset of the radiation
 driven wind, when both Roche lobes are filled as the envelope
 expands \citep{Shen2022}. 
 The designation {\it recurrent} implies that the
 outburst is observed more than once over human life timescales, 
 although all novae are thought to be
 recurrent, on long, secular timescales that vary greatly, depending
 on the mass accretion rate and the WD mass. The dependence
 on mass accretion rate is obvious;
 but it is also important to notice that, the more massive the WD
 is, the smaller is its radius, so the accumulated material is more
 degenerate and nuclear burning is ignited and becomes
 explosive with less accreted mass \citep{Yaron2005,
 Starrfield2012, Wolf2013}. Thus, the frequently
 erupting recurrent novae are more likely to host 
 massive WDs.

 RS Oph  hosts 
 an M0-2 III mass donor \citep{Dobrzycka1996, Anupama1999} in a
binary with a 453.6 day orbital period \citep{Brandi2009}.
\citet{Brandi2009, Miko2017} have presented
 evidence that the WD is  very massive, in the 1.2-1.4 M$_\odot$
 range. The effective temperature
 estimated in the supersoft X-ray phase by \citet{Nelson2008}
 in 2006 was about 800,000 K, which is indicative of a mass
 of at least 1.2 M$_\odot$.
 This implies that the WD must have grown in mass
 without ejecting all the accreted material after each TNR, since the largest
 mass of newly formed CO WDs is only around 1.1 M$_\odot$ even
 for very low metallicity \citep{Meng2008}, and
 it has spurred  interest in RS Oph as a candidate
 type Ia supernova progenitor. 
 
   RS Oph was observed in outburst in 1898, 1933, 1958, 1967, 1985, 2006
 and 2021. 
  \citet{Schaefer2004} reported  
 indications that two outbursts may have been missed in  1907 and 1945 when
  RS Oph was aligned with the Sun, namely 
  dimming episodes before and after the
 non-visibility periods of those years, resembling those observed before 
 and after some outbursts in the AAVSO optical light curve of RS Oph 
\citep[see also][]{Oppenheimer1993}.

 The 2021 outburst was announced  on  August 9 2021 22:20 UT
 by \citet{Geary2021a, Geary2021b}
 at visual magnitude 4.8.
 Immediately afterwards, the nova was also detected at gamma-ray energy
 in the GeV energy range  with
 the Fermi-LAT \citep{Cheung2022}, 
 in the energy range from 10 GeV
 to tens of TeV with {\sl H.E.S.S.} \citep{HESS2022}, 
 {\sl MAGIC} \citep{MAGIC2022} and \citet{LST2025},
 and in hard X-rays with MAXI \citep{Shidatsu2021} and INTEGRAL \citep{Ferrigno2021}.

The AAVSO
 optical light curve of RS Oph in different bands, from B to I, in 2021
 was extremely similar to the AAVSO 2006 light curve.
 The maximum magnitude was V=4.8, reached after one day. The time for a decay by 2 magnitudes t$_2$
 was 7 days and the time for a decay by 3 magnitudes t$_3$ was 14 days.
 All the subsequent evolution was smooth, and by November 14 2021
 the nova was at V$\simeq$11.2, like in 2006 at the
 same post-outburst epoch.
 The early optical spectrum was described by \citet{Munari2021}
 as of ``He/N'' type, with strong Balmer, He I and N II lines. The
 emission lines had full width at half maximum 2900 km s$^{-1}$ 
 \citep{Miko2021, Munari2021b},  however
 acceleration to up to $\simeq$4700 km s$^{-1}$ was observed 
 for some time after
 the first two days. P-Cyg profiles appeared  soon after the outburst
  in the lines of hydrogen, Fe II, O I,
 and Mg II \citep{Miko2021}; these profiles disappeared within a few 
 days.   An additional
  narrow emission component disappeared within the first few days,
 while another narrow absorption component persisted for 
 longer \citep{Luna2021, Shore2021}.
 The velocities of the lines indicated deceleration \citep{Munari2021b}
 a few days after the initial acceleration.
 Intrinsic linear optical polarization was observed $\sim$1.9 days after outburst \citep{Nikolov2023} while satellite components were detected in the optical spectra after two weeks
 in H$\alpha$ and H$\beta$, already suggesting a bipolar outflow as observed
 in the radio in 2006 \citep{Rupen2008}. High ionization lines
 appeared at about day 18 of the outburst \citep{Shore2021}. A summary
 and visual illustration of  the optical spectral changes in the
 first 3 weeks after maximum can be found in \citet{Munari2021c}.
 The radio flux, measured  in different wavelength ranges, peaked on day 10 and appeared to be non-thermal 
 since the beginning \citep{Munari2022, deRuiter2023, Lico2024, Nayana2024}. 
 Interferometric radio measurements showed a central and compact core and two
 asymmetric, elongated bipolar outflows at opposite sides of the core,
 expanding in the East-West direction at a projected velocity of 7000 km s$^{-1}$ \citep{Lico2024}.  
 
\section{RS Oph and the path to understanding
  the shocks}
 The outflow velocities measured in novae from UV, optical and
 IR spectra are of the order of thousands of km s$^{-1}$,
 and the post-shock temperature T$_{sh}$ of a thermal bremsstrahlung
 spectrum is related to the shock velocity $v_s$ (relative velocity
between the shock and the unshocked ejecta) 
 as $$ kT_{sh} \approx 1.2 \times \left( \frac {v_s} {1000~\rm km/s}\right)^2 {\rm keV} $$ 
 
\noindent \citep[see, for instance][]{Fang2020, Metzger2025}. Therefore, the plasma temperature is in the keV range, at least before
 significant radiative cooling takes place.

 The occurrence of shocks in RS Oph was first inferred {\it from the optical
 spectra}, already after the 1967 outburst and long before
 X-ray monitoring was possible. \citet{Pottasch1967}
 discussed how the optical emission lines, with their vast
 range of ionization states, could only be produced
 if the ejected flow was interacting with pre-existing
 circumstellar material of the red giant.  \citet{Gorbatskii1972} analyzed the
 nature of the coronal lines of Fe [X] and Fe [XIV] 
 and first suggested {\it that in these novae a strong shock
 propagates across the circumstellar envelope created by the red giant
 wind}. He  concluded that the shock temperature exceeded 10$^7$ K, and
 that the emergence of these coronal lines only about 3-4
 weeks after the optical maximum was due to the lack of equilibrium
 between electron and ion temperature, because equilibrium is most likely 
 reached only after at least 3 weeks, given the 
 {\it ``low ionization in the circumstellar envelope,
and accordingly by the heavy energy losses of the electron gas
 to atomic excitation and ionization''}. 
 Also Wallerstein and Cassinelli examined the evolution of the fluxes
 and flux ratios of the two coronal lines mentioned above 
  in the 1967 outburst; they presented the idea of
 shock ionization at conferences,
 but the final paper was not accepted by a refereed journal, because 
 the reviewer casted  doubts on the occurrence of shocks
 (Joe Cassinelli 2025, private communication).   
 
 In the successive outburst of RS Oph 18 years later (1985),
 more evidence of shocks became evident.
 \citet{Taylor1989} examined {\sl Very Long Baseline
 Interferometer} (VLBI) observations performed
 40 days after the optical maximum, finding two components: 
 a rapidly evolving low-frequency non-thermal one
 and a high-frequency thermal component that
 reached peak flux density much later than the non-thermal emission. 
 \citet{Bode1985} examined the first detection of X-rays, done
 with {\sl Exosat},  55 days after 
 maximum, concluding that shocks definitely did occur
 and proposing they follow the same evolution as in supernova remnants: 
  the Sedov phase is reached when the mass of the ejecta
 is about equivalent to the mass of the plowed red giant wind,
 the plasma temperature T decreases
  with time as T $\propto t^{-2/3}$, and the X-ray luminosity
 as  L$_{\rm x} \propto t^{-1/3}$. 
 A ``phase III'' follows with  radiative
 cooling, so
 that temperature and X-ray luminosity decrease with time as T$\propto$ t$^{-1}$
 and L$_{\rm x} \propto$ t$^{-1.5}$, respectively.  
 On the basis of the sparse
 data at the time, these authors assumed that the outflowing plasma was still in the Sedov phase on day 55,
 at the time of the {\sl Exosat} observation,
 but it was transitioning to ``phase III''.
   A fit with a bremsstrahlung model indicated a post-shock plasma temperature of 0.75 keV, although the authors noted that assuming that most of the flux
  below 2 keV
  must have been in unresolved emission lines, the correct value would have been 
 of about 0.25 keV. Assuming the distance of 2.4 kpc generally accepted today \citep[see discussion by][]{Orio2023} the absorbed luminosity was about 
 2 $\times 10^{35}$ erg s$^{-1}$. 
 Unfortunately, it is almost certain that having only broad band measurements as late as day 55 led to mistaken estimates
 due to contamination by the luminous SSS,  the major contributor to the X-ray flux 
 at that epoch in the more recent outbursts. 
 The huge soft X-ray flux of the SSS in RS Oph  in the 0.2-0.8 keV
 range after the
 first month was not known; it was first measured only in 2006.

 However we note that, regardless of this source of confusion, the shocked plasma of RS Oph 
indeed emits X-ray flux for a long time:  \citet{Contini1995}
 modeled lines in the optical spectrum as due to shocked material
 as late as on day
 201 of the outburst, presumably after the disappearance
 of the SSS (in the following eruptions monitored in X-rays,
 the thermonuclear burning turned
 off completely by day 200).  Assuming that the ejecta
 did not slow down much, the outer shell at this post-outburst
 epoch should have been at a distance of approximately
 5 $\times 10^{15}$ cm from the 
WD,  so the calculations by \citet{Contini1995} suggested
 that  the second X-ray detection done with {\it Exosat} 250
 days after the maximum of the 1985 outburst was due to shocks,
 while the  supersoft X-ray source had already faded. This was confirmed 
 to be the likely case when in 2006 an emission line
 spectrum with no measurable continuum was observed with {\it XMM-Newton}
 239 days after the outburst, with an absorbed X-ray flux  in
 the 0.2-1.5 keV range of 
$\simeq 9 \times 10^{-13}$ erg s$^{-1}$, still quite higher than in 
quiescence \citep{Nelson2008}. 
  Furthermore, we know today that
 shocks still occur, even in short-orbital-period novae, 
 3 to 6  months after
 maximum \citep[see the cases of YZ Ret and V1716 Sco;][]{Mitrani2024, Mitrani2025}. 

 \subsection{The first intensive X-ray monitoring in 2006} 
 The ``history'' of RS Oph teaches that
 understanding the development of the shocks in novae 
 occurring in symbiotic binaries requires 
 frequent observations in different wavelength ranges. 
 As  more and better-quality observational material
 was gathered, the models were gradually revised.
 The 2006 outburst was the first one of this nova well monitored in X-rays: 
 with {\sl RXTE} in an initial ``hard'' phase, then daily
 with {\sl Swift} for the whole duration of the outburst 
\citep{Bode2006, Hachisu2007, Osborne2011,Sokoloski2006}.
The nova was also 
observed at several epochs with high spectral resolution with the gratings of
 {\sl Chandra} and {\sl XMM-Newton} \citep{Ness2007,
 Nelson2008, Drake2009, Ness2009}. It was
 thus understood that shocks first occur {\it very early in the outburst}.  

 By fitting the X-ray spectrum observed with {\sl RXTE} during the first three weeks in the 2-25 keV range  
  with a bremsstrahlung model,  \citet{Sokoloski2006}
 suggested that  
 {\it the plasma temperature} decreased as t$^{-2/3}$, as expected
 {\it already in the Sedov phase},  implying that 
  the  shock velocity decreased as t$^{-1/3}$.
  The rate of decay of the {\it X-ray flux}
 in the 2-25 keV range of {\sl RXTE}
 seemed to indicate that the expansion was not spherically symmetric.
 The authors inferred
 that by the day 1.7, time of the first exposure,
 the nova outflow had swept up about 10$^{-7}$ M$_\odot$ of
 material (assuming a circumstellar medium - hereafter CSM - with a
 particle  density of 
 10$^9$ cm$^{-3}$).  These authors assumed that the mass ejection was 
 episodic and non-continuous (unlike in the nova models)
 and reasoned that the estimated
 swept up material  is approximately also the total
 mass of the ejecta: the nova models predict such a low ejected
 mass only for a near-Chandrasekhar mass WD, raising interest in RS Oph
 as a possible type Ia SN progenitor.

   {\it These conclusions were  based on 
 data  with incomplete energy coverage at low energy and they were
 revised}  by \citet{Bode2006} thanks to {\sl Swift} 
  X-Ray Telescope (XRT) monitoring in the 0.2-10 keV range during 
the whole outburst, and to  the Burst Alert Monitor (BAT) 
monitoring in the 14-25 keV range during the first 6 days.  
  The X-ray flux in the energy range
 covered with {\sl Swift} increased until day 5,  
 and  it was measured to have decreased only in the next exposure, done
 on day 8.  \citet{Bode2006}
 inferred the presence of prominent, albeit mostly unresolved,
emission lines in the X-ray spectrum, so
  they fitted the broad band spectrum with a thermal plasma
  model that included lines ({\sl MEKAL} in XSPEC). After day 5, the rate
 of decrease of the velocity of the shocked material
 and of {\it unabsorbed} flux 
  F$_{\rm x}$ followed trends with time as v$_s \propto$ t$^{-0.6}$ and  
 F$_{\rm x} \propto$ t$^{-1.5}$, respectively, as expected already in phase III,
 implying  very efficient radiative cooling.
 The fit also yielded that the column density decreased as 
  N(H) $\propto$ t$^{-1/2}$, which is also consistent with ``phase III''. 
 In the next section, we describe how these
 conclusions were  once again
revised when \citet{Page2022} re-analyzed the 2006 data
 with two temperature components, after examining the 2021 X-ray lightcurve
 and the results of high resolution X-ray spectroscopy. 
 
 Although the deviations from
 spherical symmetry could not be inferred only from
 the broad-band X-ray data and 
 the  ejected mass may have been more than triple the
 estimate of \citet{Sokoloski2006}, 
 lack of spherical symmetry
was indeed measured later at radio wavelengths. The early observations
 with the VLBI revealed 
 in fact mostly non-thermal radio flux, likely starting on day 3.8, and 
 initially expanding more on one side \citep{OBrien2006}. This was 
 interpreted as a bipolar outflow and 
 asymmetric expansion was observed in the radio imaging of the
 2021 outburst \citep{Lico2024}.

\subsection{The high resolution X-ray spectra of 2006}
 For rigorous measurements and models of the X-ray spectrum,
 high spectral resolution X-ray data are needed. In 2006,
   X-ray grating exposures started only on day 13.8, 
 almost simultaneously with the {\sl Chandra} high energy 
 transmission gratings (HETG, wavelength range 1.2-21 \AA \ or energy range 0.4-10 keV) and with the {\sl XMM-Newton} reflection grating
 spectrographs (RGS, wavelength range 5-30 \AA \ or energy range 0.3-2.5 keV). 
 A thermal model of shock
 ionization was adopted in all the papers that addressed these impressive
 data \citep{Nelson2008, Ness2009, 
 Drake2009}. \citet{Nelson2008} noted how the lines
 appeared blueshifted by an amount that was dependent
 on wavelength and ionization state and \citet{Drake2009} explained this
 as an artifact of differential absorption in both the red giant wind and in
 the ejected material. The red wing of the line (formed
 in the receding ejecta) is more ``corroded'' by absorption, which
 of course is more effective at long wavelengths, resulting
 in an apparent blueshift varying with wavelength. From a detailed
 analysis of the line profiles, \citet{Drake2009} also suggested collimation
 of the ejecta in the direction perpendicular to the line of sight,
 supporting the bipolar outflow model inferred from the radio data. 

The  emission lines were produced
 in transitions with a wide range of ionization potentials.
 \citet{Nelson2008} examined the line ratios of the He-like triplets in 
 the spectra of day 13.8 and fitted the spectra
 with a super-position of regions of plasma at different temperature
 (respectively 16.84, 2.31, 0.92, and 0.64 keV, with 
 absorbing column of 1.2 $\times 10^{22}$ cm$^{-2}$ and assuming
 solar abundances), with all the plasma already in collisional 
 ionization equilibrium (CIE). 
 A different approach was taken by \citet{Ness2009}, adopting 
 non-solar abundances as additional parameters to fit
 the spectrum of day 13.8 and a late one taken on day 111.
 For day 13.8 they obtained a fit with three zones at temperatures
 of 4.19, 0.74 and 0.30 keV, respectively, 
 and enhanced abundances relative to solar values 
 by factors from 2 to 7 for nitrogen, while only iron resulted  depleted.
  \citet{Ness2009}
 also estimated the physical parameters of the plasma with an
  alternative {\it emission measure distribution} method,
  confirming only the overabundant nitrogen and depleted iron,
 while the other elements resulted only slightly depleted with respect to the
 solar value. This  method yielded two peaks of plasma temperature, corresponding
 to 0.17 keV and 0.8 keV, respectively,
 and only an upper limit for the highest temperature 
 from the iron lines: there was significant flux in the Fe XXV
 lines, while Fe XXVI was not even detected, which was not consistent with
 the derived range of temperatures. 
 While other line ratios indicated that the plasma was in CIE, equilibrium
 had not be reached yet for the transition from
 He-like to H-like iron, consistently with a ionization energy of 8.8 keV,
 higher than the highest plasma temperature of 4.19 keV
 derived from the best fit. The conclusion of \citet{Ness2009} was that  
 by the 14th day of the outburst 
 atomic transitions already occurred in equilibrium,
 except the ones involving hydrogen-like and helium-like iron. 

 As the plasma cooled, new high resolution spectroscopic
 X-ray observation 
 were done on day 26 with the {\sl XMM-Newton} RGS. An emission line spectrum was still measured,
 but a flare occurred during which  elevated SSS continuum
 appeared. Simultaneously with the continuum flare,
 a quasi periodic oscillation with
 an approximate period of 35 s was detected for the first time. This 
 modulation was attributed to the central source (the WD)
 whose photospheric radius had  by this time
 receded during the constant bolometric
 luminosity phase described above, as the atmospheric temperature increased.
 The modulation was  almost 
 always detectable in the  flux due to the SSS, namely
 below 0.8 keV, which became the dominant source 
 with almost all flux emitted below 0.8 keV.   During the flare 
 of day 26, a puzzling new set of transient
 emission lines emerged at low energy.
 Identifications were proposed by \citet{Nelson2008}  assuming  
 that the transient lines were due to material
 escaping at high velocity, with a
   blue shift of at least 8,000 km s$^{-1}$, close to the escape velocity from the WD.
 A possible interpretation was that of a new episode of mass
 ejection occurred at much higher velocity than in the
 initial wind. \citet{Nelson2008}
 did not attempt to  fit these spectra, but noted that the H-like/He-like
 ratio of the Mg lines indicated a temperature of about 8 $\times 10^6$ K
 ($\simeq$0.7 keV). The 
 transient additional spectrum is the topic of a paper in
  submission phase by Mitrani et al., 2026,  who
 found a possible
 origin in ``clumps'' of freshly synthesized material, brought to the surface 
by convection. 

 \citet{Ness2009} also
 fitted X-ray spectra taken of the late
 phase after the SSS decline. They tried subtracting an
 ``approximate'' WD continuum to fit the residual emission
 line spectrum during the SSS phase.  Their conclusions 
 can be summarized as follows:
 
  a) At least three APEC components 
 were needed to explain the spectra for both days
  13 and 26;
  
   b) The hottest component contributing to
 the X-ray flux cooled very rapidly,
 down to a value kT$_{\rm max} \leq$1.8 keV on day 26. Later,
 on days 112, 206 and 239.2, they obtained  kT$_{\rm max} \leq$ 0.78 keV;  
 
 d) The ``softest'' APEC plasma component(s) in X-ray flux below 1 keV, still
 detected after the SSS faded,
 cooled instead very slowly and remained almost unchanged during the 
 first weeks. 
\begin{table}
\begin{center}
\begin{tabular}{ccccccc}
\hline
 & F(Day 1-2) & F(Day 4-5) & F(Day 9) & $\alpha(F_{\rm abs})$ & $\alpha(F_{\rm un})$ & $\alpha(T(high))$ \\
 & erg cm$^{-2}$ s$^{-1}$ & erg cm$^{-2}$ s$^{-1}$ & erg cm$^{-2}$ s$^{-1}$ &  &  & \\
\hline
 {\sl H.E.S.S.} & 10$^{-12}$ & 8 $\times 10^{-12}$ & $\approx 2 \times 10^{-12}$ &  -- & 1.3/1.4 & -- \\
 {\sl Fermi} & 5 $\times 10^{-9}$ & 1.7 $\times 10^{-9}$ & 7 $\times 10^{-10}$ & -- & 1.3/1.4 & -- \\ 
 {\sl NICER \citep{Orio2023}} & 7.5 $\times 10^{-10}$ & 1.36 $\times 10^{-9}$ & 7.7 $\times 10^{-10}$ & $\simeq$1 & $\simeq$1 & 0.6 \\
 {\sl NICER \citep{Islam2024}}            &                       &                       &                       & $\simeq$1 & 1.2$\pm$0.1 & 1.05 \\
 {\sl Swift} & 7.3 $\times 10^{-10}$ &  1.27 $\times 10^{-9}$ & $\geq 6.2 \times 10^{-10}$ & 0.9/1 & -- & 1.18 \\
 {\sl Swift 2006} & --               & 2.0 $\times 10^{-9}$ & -- & ---  & 1.5 & 0.77 \\
 {\sl NuSTAR}     & --               & --                & 6.4$\pm$0.3 $\times 10^{-9}$  & 1 & -- &   1.4 \\
 {\sl NuSTAR+NICER} & --             & --                   & 8.5$\pm$0.4   $\times 10^{-9}$ 
  & -- & -- & -- \\
\hline
\end{tabular}
\caption{The flux measured on different post-outburst days with
 the X-ray and gamma-ray instruments, and the slope $\alpha$ indicating
 the trend with time (where 
 the decay with time $t$ follows a law $\propto t^{- \alpha}$) of 
 measured flux, unabsorbed flux at the source estimated from the spectral fit,
 and plasma highest temperature in these fits. For {\sl NICER}
 we report the power law index $\alpha$
 estimated for the same data for unabsorbed flux
 and temperature in the two different works,
 \cite{Orio2023, Islam2024}.}

 \label{tab:1}
\end{center}
\end{table}

\section{The gamma-rays in 2021}
 Table \ref{tab:1} summarizes the X-ray and gamma-ray fluxes that were measured with the different X-ray missions
 in 2006 and 2021 at early epochs, and the rate of decay in the first
 three weeks by assuming a power law to model the light curves.  

In this section, we briefly review the gamma-ray monitoring,
 while the X-ray results of 2021 are described in the next section.
By the time of the last outburst of RS Oph, the detection of gamma-rays
 with {\sl Fermi Large Area Telescope (LAT)} was already expected.
However, RS Oph is the first and only nova from which gamma-ray flux was
 also detected with the Cherenkov telescopes, at very high energy
 (several TeV) \citep{HESS2022, MAGIC2022},
 as it had been previously predicted by \citet{Tati2007} and \citet{Metzger2015}.
 Already in 2007, \citet{Tati2007}
 calculated that  the particle acceleration rate
 (fraction of shocked protons that are subject
 to diffusive acceleration) must be $\eta_{\rm inj} \geq 10^{-4}$.
 In the years following this theoretical paper, 
 {\sl Fermi} observations demonstrated 
 that generally, novae are indeed gamma-ray sources early in the outburst,
 because  over a dozen other  novae since 2010 were detected
  \citep[see][and numerous Astronomer's telegrams]{Cheung2016, Franck2018, 
 Sokolovsky2022, Sokolovsky2023}. 

 The gamma-ray emission is due to particle acceleration. 
In novae  in symbiotic binaries, such as RS Oph, 
it is likely of hadronic origin, namely due to the decay of pions produced in 
interactions of accelerated protons or ions with the ambient material 
 \citep[see][and references therein]{Diesing2023}. 
 These novae have a dense red-giant wind, mostly accumulated
 in spiral structures in the orbital plane; the nova super-wind
 collides with the red giant
 wind, which is dense and slow, with 
 velocity of tens of km s$^{-1}$ \citep[see discussions by][]{Cheung2022, Diesing2023}.

 Powerful shocks have been observed in X-rays in
 many novae \citep[e.g.][]
 {Balman1998, Drake2010, Orlando2012, Orio2012} 
 but it is puzzling that the peak 
 of the X-ray flux was not  simultaneously observed in gamma and X-rays, 
 even if the outflow velocity 
 measured in optical spectra is  only compatible with
 X-ray emission.  Before the last
  outburst of RS Oph,  not only the X-rays were often initially detected 
 in novae several days after the gamma-ray peaked, but also the X-ray flux 
 was always orders of magnitude too small to be associated with the
 observed gamma-rays. Instead, peaks of
 optical brightness were observed simultaneously with
 the maxima of gamma-ray flux or with a delay of about a day,
 so \citet{Aydi2020}
 suggested that the X-rays are reprocessed into optical light.
 This implies that {\it the rate-of-decay vs. peak magnitude relation
 is  much less significant than previously thought, since in
 some novae there may be a large
 contribution to the optical flux due to reprocessed emission from 
 the shocks}. 
  
 The X-ray flux may be heavily absorbed
 by large column density, but another explanation
 of this phenomenon  has recently been given by \citet{Metzger2025},
  if in the CSM there is a  much cooler gas than the material ejected
 from the WD.  This  model was studied to explain some
 short period  novae for which
   several radio and some optical observations are indicative of a slow,
  dense toroidal outflow and a faster, less dense poloidal outflow 
\citep[see examples in][]{Chomiuk2021}. 
 Turbulence efficiently balances the shock heating by mixing reducing the volume of hot gas, suppressing the X-ray luminosity. 
There is indeed evidence of mixing between shock-heated plasma and cold gas in the area surrounding novae at late stages in the outburst \citep{Mitrani2024, Mitrani2025}.
 In novae  occurring in symbiotic binaries, the slow toroidal flow
 should not be emitted  because it is most likely due
 to the embedding of the secondary in a common envelope during
 the early phase of the outburst. 
 With such large orbital separation, radiation pressure causes a fast outflow 
 before the outflow reaches the red giant \citep{Shen2022}. 

 Table 1 shows that for RS Oph, in both GeV and TeV energy ranges there was an initial rise already after a day, but while the GeV luminosity peaked in the second day \citep{Cheung2022}, the TeV luminosity peaked on days 4-5 \citep{HESS2022}. In both ranges, the luminosity decayed with time as t$^{-\alpha}$, where $\alpha$=1.3-1.4 \citep[see][]{Diesing2023}.
 The flux peaked in the ``softer'' range of {\sl H.E.S.S.},
 with a power law index $>$3, while a 1.9 power law index was measured for the Fermi-LAT spectrum in the 0.1-13 GeV range.
 By modeling the shocks in detail, \citet{Diesing2023}
 showed that a single shock cannot simultaneously explain RS Oph's GeV and TeV emission,  their spectral slopes
  and distinct lightcurve peaks.  Two or more shocks reproduce instead
 the observed gamma-ray spectrum and temporal evolution.
 If TeV and GeV range gamma-rays arise in the same shock, the TeV flux
 should be larger, and clearly this was not the case of RS Oph.
 An alternative explanation has also been proposed, namely that interactions
 between gamma rays and optical photons emitted in the outburst
 initially caused differential absorption, so that the medium was 
 opaque to photons with energy above 200 GeV only in the first days of 
 the outburst \citep{Phan2025}. 

\section{New X-ray monitoring in the 2021 outburst} 
 In X-rays, the 2021 outburst
  was monitored with the {\sl Swift} X-Ray Telescope
 since 0.37 days after the optical peak, and with
 the Neutron Star Interior Composition Explorer camera
 ({\sl NICER}) aboard the International Space Station
 (ISS) since day 1.27. Both instruments offer only  
 short snapshots, that in most cases are not longer than 1000 s. 
 Each of the {\sl Swift} and {\sl NICER} databases comprise more than 200 
 exposures between the first day of the outburst and the beginning of
 2021 November, when observations had to be
 interrupted because the nova was too close to the Sun. 
 
 Several conclusions were derived from this monitoring, and one
 lesson learned is that more frequent X-ray high spectral resolution
 spectra 
 are needed to constrain the cooling curve and the trend of
 the physical parameters, because the results of the model
 fits to broad band data often are not unique.    
 The {\sl Swift} data analysis was done by \cite{Page2022};
  \citet{Orio2023} analyzed the {\sl NICER} data, which for 
  the first month of the outburst were also
  revisited by \citet{Islam2024} with a different approach 
 in the spectral fits.
 Although the evolution of the X-ray lightcurve after the
 rise of the SSS continuum was quite different from 2006, 
 but the lightcurve of the first month
 was very similar to 2006. 

 The hottest
  component  in the fits by \citet{Page2022, Orio2023},  or the only component
  for \citet{Islam2024}, cooled very rapidly like in 2006.   
 As the WD became a luminous SSS at the end of the first month,
 there was still significant emission arising from the 
 shocked ejecta.
The {\sl Swift} XRT spectra were fitted with a
CIE (Collisional Ionization Equilibrium) model with two temperature
 components by  \citet{Page2022}, who revisited also
 the 2006 broad band spectra of the {\sl XRT}. These authors found that 
 also the 2006 spectra were best modeled
 with two plasma components, like the 2021 ones.
 The highest temperature could not be well constrained on days 3 and 5, 
 but the flux clearly increased on day 5.
 The table in \citet{Page2022} does
 not report the value of the {\it unabsorbed} flux;
 however, applying their model we find that the flux
 decay rate as $t^{-1.5}$ suggested by \citet{Bode2006}
 was not confirmed; in fact the new fits yield different column density
  and there
 is interplay with the much less hot thermal component.
The best fit parameters in the table in the paper of \cite{Page2022}
 imply a peak unbsorbed flux 5.29 $\times 10^{-9}$ erg s$^{-1}$ in 2021,
 and about 4.59 $\times 10^{-9}$ erg s$^{-1}$ in 2006 \citep[more
 than double the estimate of][]{Bode2006}.

 {\sl NICER} measured several emission lines 
 above 0.5 keV: at least in the first week 
 the spectrum was  more complex than a CIE plasma because collisional
 ionization equilibrium was not reached. 
 {\sl NICER} monitored the nova densely already during the first 5 days,
 allowing to follow the flux rise 
 during the first 4 days, followed a plateau in the fifth day,
 largely due to the decrease in column density N(H) \citep{Orio2023, Islam2024}.
{\it The initial column density was much higher than evaluated in 2006,
 resulting in higher estimates for the unabsorbed flux}, in agreement also with the new models for the {\sl Swift} data.

Before day 5, \citet{Orio2022} modeled the broad band spectra  
finding that the unabsorbed flux peaked on day 5.67 like
 the absorbed flux that wa sactually measured; however in their
 model the emission measure increased while the temperature
 was already cooling after a peak on day 1.9 at kT=25.73$\pm$1.50 keV. 
 Both
\citet{Orio2022} with {\sl NICER} and  \citet{Page2022} with {\sl Swift}, found that
 a second, cooler component of CIE plasma was necessary to fit all spectra
 from day 2, initially with a temperature of only $\simeq$0.2 keV, 
and from the 10th day, between 0.55 and 0.9 keV. 
A third soft component, a blackbody, was added on day 20, to model
 low-level supersoft flux even before the SSS continuum clearly emerged. 
 The inclusion of a partially covering absorber for
 the {\sl NICER} spectra results in a different cooling rate of the ``hotter'' 
 component than in the {\sl Swift} analysis, as shown in Fig. 1.

 \citet{Orio2023, Islam2024} analysed the same {\sl NICER} data,
 but differ in estimating the time of onset of collisional
 ionization equilibrium.  Several emission lines, down to the  
 lowest energy transitions of the (blended)  \ion{N}{6} He-like
 complex around 0.43 keV, were
 measurable. The total flux in the He-like triplets
 was unusually high compared to that in lines due H-like transitions in
 CIE, suggesting  that the plasma was not in  equilibrium. However, fits with
 non-equilibrium models require more free parameters; thus a good fit 
 for these spectra is not reached with a unique model.
 \citet{Orio2023} attributed the flux increase in the first
 5 days to decreasing column density of absorbing material, 
  but they found that the temperature already decreased between day 1.9
 and day 5.67.  
 For these authors,  only the spectra
 of the first 6 days should be fitted with a non-equilibrium model of 
 thermal emission and the 
 addition of a partially covering absorber, while for
 \citet{Islam2024} non equilibrium persisted until day 18.
 Given that the electron temperature is directly
 dependent on the square of the velocity as
 kT$\simeq$1.2 keV (v/(1000 km/s)$^2$, this yields a velocity
 of 4630 km s$^{-1}$, very close to the estimates form the
 optical spectra. 
 \citet{Islam2024}, instead, reasoned that the peak of the flux and of
 the plasma temperature should coincide, so their model is consistent 
 with a parallel rise of flux and temperature until the observation
 on day 5.67.
  
\begin{figure}
\begin{center}
\includegraphics[width=130mm]{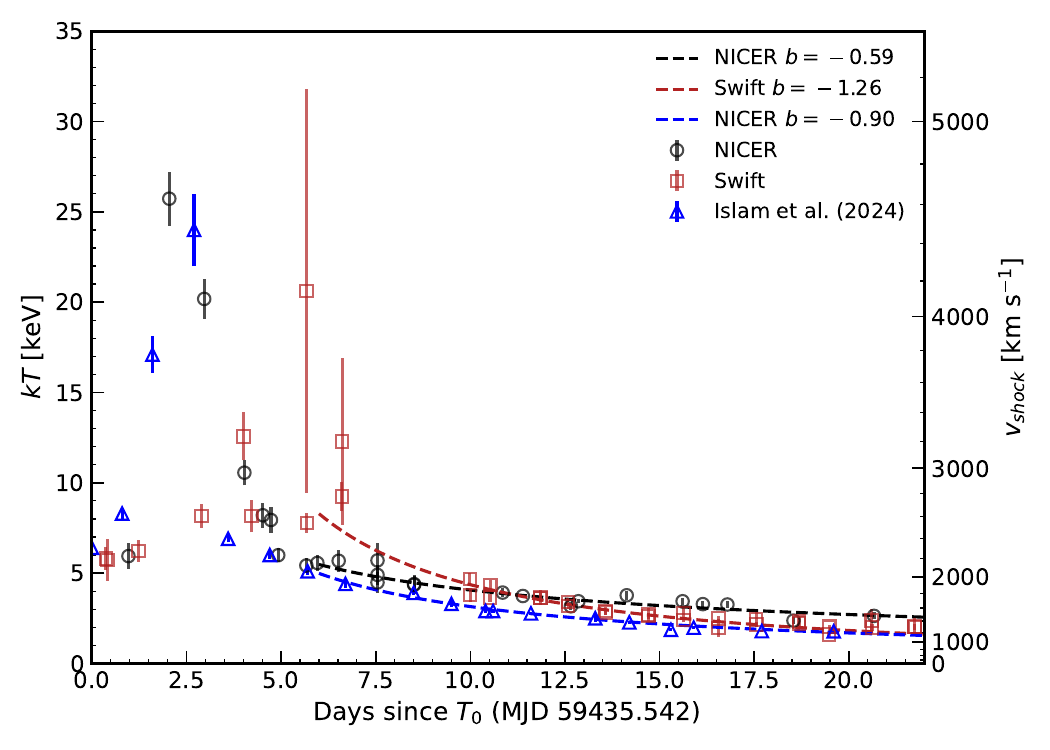}
\end{center}
\caption{The time evolution of the highest temperature 
 obtained in the spectral fits by \citet{Page2022, Orio2023} and of the
 temperature in \citet{Islam2024}. From day 5, when the flux
 declined, a different trend in time - as shown in the inset where
 the slope is indicated as $b$ - was derived in  each of the three papers, 
depending on the model adopted for the fit.
 }
\label{fig:1}
\end{figure}

  From day 8, both \citet{Orio2023} fitting the  
 {\sl NICER} spectra and \citet{Page2022} fitting the {\sl Swift} spectra
  obtained the best fits with two-components of plasma in CIE.
 However, while
 the hotter component was found to cool differently - as shown in Fig. 1 -
 with the two models, respectively as $t^{-0.59}$ for \citet{Orio2023}
 and as $t^{-1.26}$ for \citet{Page2022},
 the ``cooler'' one had an almost constant temperature close to 0.9 keV,
 without significant cooling.
 \citet{Islam2024} used instead a non-equilibrium  model
 with four components, all at the same temperature, but with
 different ionization timescales, until day 18. With this assumption, the
  ``cooler'' component was not needed, but - like in \citet{Orio2023} - an 
 additional partially covering absorber was necessary for the fit.
 These authors derived kT=8.3$^{+0.3}_{-0.4}$ on day 2.7, and 
 a peak at kT=24$^{+1}_{-3}$ keV on day 4.6 \citep[about the same value of the maximum temperature found by][]{Orio2023}. 
 This results in a different trend of the plasma temperature with time then found in the
 two previous paper,
  as shown in Fig. \ref{fig:1}: a slope with a -0.90 exponential factor,
 closer to the result obtained by \citet{Page2022}
 assuming equilibrium and no partially covering absorber.

\citet{Islam2024} found that the emission measure 
 (hence the absolute flux) decreased as t$^{-1.2\pm0.1}$.
 We notice that the ionization time scales in the work
 of \citet{Islam2024},
 n$_e \times t$ of 10$^9$, 10$^{10}$, 10$^{11}$ and 10$^{12}$
 s $\times$ cm$^{-3}$ imply, on average,  very low electron density 
 where the shocks occur {\it if the shocks occurred at the
 beginning of the outburst}. The two shorter ionization time
 scales in this model dominate after two weeks
 (t$\geq 1.2 \times 10^6$ s). 
 Assuming the time elapsed since the beginning of the outburst, this
 implies electron density as low as n$_e = 10^{3}$ cm$^{-3}$,
 which appears very unrealistic for RS Oph. 
\citet{Lico2024} found electron
 density of 10$^{7}$ cm$^{-3}$ on day 15, which is consistent 
 also with other estimates for RS Oph and for many other novae.  
 Although the electron density decreased quite
 steeply in time, \citet{Munari2022} still derived
 an average n$_e = 10^6$ cm$^{-3}$  34 days after the outburst. 
 In this model, only if there were episodes of mass ejection at different
 epochs, and the  plasma was  shocked
 several times during the 3 weeks period of their analysis,
 n$_e$ may have even been constant around a value of 10$^{6}$ cm$^{-3}$.
 However, in the case of several different shocks there is no reason
 to assume a constant temperature, and the actual physical development 
 may be much more complex. 

 Although all authors agree that in the first four days 
the column density was close to, or exceeded, a value of 10$^{23}$ cm$^{-2}$,
 the interplay of plasma temperature and N(H), and the fits with or without a partially covering absorber yield greatly different model-dependent
 values for the {\it peak absorbed flux}. 
 \citet{Islam2024}
 estimated about 4 $\times 10^{-9}$ erg s$^{-1}$ cm$^{-2}$,
 while \citet{Orio2023} inferred
 a value of  2 $\times 10^{-8}$ erg s$^{-1}$ cm$^{-2}$. 
 
 High resolution X-ray spectra measured since the end of the 3rd week
 (days 18 and 23
 after optical maximum) with the grating spectrographs of {\sl Chandra} and {\sl XMM-Newton} confirmed that the ``hotter'' plasma component cooled rapidly,
 from kT$\approx$25 keV at maximum inferred by the above authors,
 down to kT=2.4 keV on day 21 \citep{Orio2022}. At this stage, 
 two components of plasma in collisional
 ionization equilibrium fitted the spectra on day 18 and three
 components were necessary on day 23. 
Both in 2006 and in 2021, derivation of velocity from
 the grating spectra is hindered by 
 the  different broadening of each emission
 line, that  seemed to vary for the same line even within few hours 
 \citep{Nelson2008, Ness2009, Orio2022}.
\section{NuSTAR data: the cooling thermal spectrum of day 9
 and upper limits to the non-thermal flux}
On 2021 August 17, we obtained a long Director Discretionary Time observation with {\sl NuSTAR} in the 3-79 keV range. 
 A preliminary analysis of these data was reported in an 
 Astronomer's Telegram
\citep{Luna2021} and in conference proceedings \citep{Orio2025}. 
  They add important information to
 interpret the shock evolution. {\sl NuSTAR} is the ideal instrument to determine the exact value of the peak temperature and to measure the precise value of the flux in a  much higher energy range than {\sl Swift} and {\sl NICER}. Unfortunately, in 2006 it had not been launched yet and in 2021 it was impossible to schedule the exposure earlier.  

Since {\sl NuSTAR} is in a low Earth orbit, 
 with the exception of small continuous viewing zones near the
orbit poles, most of the sky is occulted by Earth for half of
 its
orbit. Unlike {\sl Swift}, NuSTAR cannot repoint multiple times per orbit 
and needs to  wait for the
target to reappear from behind the Earth limb (it
 cannot observe during the 
occultations). Having a 40ks 
exposure spread over more than a day is the standard, and usually the
only available observing mode.
 In our case, the exposure was carried out over 15 satellite orbits with
 an effective  observation time of 39.5 ks over almost 1.2 days, during which the count rate decreased by approximately 30\%.   The average count rate in the 3-79 keV  NuSTAR range was 12.95$\pm$0.02 cts s$^{-1}$ for the FPMA detector and 12.03$\pm$0.02 cts s$^{-1}$ for the FPMB detector. 
There were almost no counts above the background above 30 keV. 

The light curve in the 3-30 keV range is shown
in Fig. 2: it shows a clear trend as t$^{-1}$.
We modeled the  spectra obtained during each satellite orbit in
 the 3-30 keV range with one
 APEC component and found  that the temperature decreased as shown in Fig. 3: the cooling followed an approximate t$^{-1.3}$ trend with time, quite steeper
 than derived by fitting the {\sl Swift} and {\sl NICER} data over
  time periods of over 2 weeks.  
While it is  possible that the rate of decrease of the temperature with time varied from day to day, it is more likely that the higher energy range 
isolated with {\sl  NuSTAR} may have allowed a more reliable measure of the trend of the highest temperature with time, 
without contamination of the cooler component.

Fig. \ref{fig_nustar_nicer_comp} shows the {\sl NuSTAR} and {\sl NICER} 
spectra of a snapshot of time during which the nova was observed with 
both instruments, on 2021 August 9 between 1:59 and 3:03  with NICER and 
with  {\sl NuSTAR} on the same day between 2:38 and 3:41.
 It was not possible to schedule both instruments simultaneously, and this is the only short exposure interval in common. The fit was done in the 0.25-10 keV range for {\sl NICER} and in the 3-25 keV range for {\sl NuSTAR}.  
We fitted the spectrum with an absorbed, velocity broadened optically thin thermal plasma (\texttt{TBabs $\times$ BVAPEC}), with parameters shown in the third column of Table \ref{tab:2} (Model 1). 
We note
 a typical feature of the broad-band X-ray spectra of
 novae in symbiotic systems, namely the (unresolved) triplet of
 Fe XXV, which distinguishes them from the more common novae in CV-type
 system, where emission lines of Fe XXV and Fe XXVI are hardly
 ever measurable. This is not only due to the generally
 higher plasma temperature, but also, and mainly, to the highly non-solar
 abundances of the ejecta when they are not mixed with the red giant wind.
 The iron features are instead well detected in RS Oph despite the sub-solar
 iron abundance noted above.
The light curve in the 3-30 keV range is shown
in Fig. 2: it shows a clear trend as t$^{-1}$.

Assuming two different ``BVAPEC'' regions at different temperature, however,
 seems to improve the fit, at least for this exposure interval.  
The addition of a  partially covering absorber with a small addition of a uniformly distributed column density was found to be necessary for the {\sl NICER} spectra by both \citet{Orio2023} and \citet{Islam2024}. 
The better signal-to-noise of {\sl NICER} drove the fit,
 so we had 
to include both the partially covering absorber and the thermal component with plasma temperature close to 1 keV.  
By fitting the two instruments together for this common time interval,
 the hotter thermal component with temperature turns out to be 3.9 keV,
 consistently, within errors, with all the results 
obtained by \citet{Page2022} for {\sl Swift} and by \citet{Orio2023, Islam2024} with {\sl NICER} immediately before and after day 9.

  In this fit also the abundances were free parameters, 
 following \citet{Ness2009},
resulting in most elements except
 iron a few times enhanced with respect to the solar value, however 
with large uncertainties. Only the
 iron abundance is constrained with a relatively small statistical error, 
 resulting to be only about a third the solar value (see Table \ref{tab:2});
 the same value was derived 
fitting the high resolution X-ray spectra  both in 
 2006 \citep{Ness2009} and in 2021 \citep{Orio2022}.
We did not bin the {\sl NICER} spectrum with the same factor as the {\sl NuSTAR} one because there are several emission lines, 
whose information would be lost with large
 spectral bins, blending them with the continuum. 
 The large residuals in Fig. \ref{fig_nustar_nicer_comp} around the iron feature is due mainly to calibration uncertainty of the energy of this 
 line in {\sl NuSTAR}, but it agrees with the Gaussian
 centroid found with {\sl NICER} within $\approx 2 \sigma$.   

We also searched a best fit for the {\sl total} spectrum of all 
 coadded {\sl NuSTAR} short exposures and 
 include it in the first column of Table \ref{tab:2}. It  required two ``hot''
 {\sl BVAPEC} components, at 2.57$\pm$0.14 keV and 6.16$\pm$0.27 keV,
 respectively, with only a partially covering absorber.  This
 may be the result of  adding short spectral exposures a plasma
that was slowly cooling during the 26 hours of the exposure. However, 
 when  applied to the single continuous
 exposure intervals, as shown by Model 2 in Table 
 \ref{tab:2}, the assumption of two
 zones at different temperature gives a marginal improvement.
 These two zones would be in addition to the softest component
 around 0.9 keV.

\begin{table}
\begin{center}
\begin{tabular}{cccccc}
\hline
  & {\sl NuSTAR} & {\sl NuSTAR}  & {\sl NuSTAR}+{\sl NICER} & {\sl NuSTAR}+{\sl NICER} & All day {\sl NuSTAR}  \\
  & 1 comp.& 2 comp. & Model 1 & Model 2   & Average \\
  & 3-25 keV & 3-25 keV & 0.2-25 keV  & 0.2-25 keV  & 3-25 keV \\
\hline
$\chi^2$/d.o.f. & 1.5 & 1.1  & 1.1 & 1.1 & 1.5 \\
 N(H)$_p$ $\times 10^{22}$ cm$^{-2}$ & -- & -- &  5.20$\pm$0.20 & 5.28$\pm$0.29 & 22.19$\pm$5.9  \\
  Cov.F. & -- &  -- & -- & 0.71$\pm$0.01 & 0.44$^{+0.11}_{-0.07}$  \\
N(H)  $\times 10^{22}$ cm$^{-2}$ &  1.20$^{+1.14}_{-1.19}$  & 4.59$^{+1.64}_{-1.40}$ & 0.56$\pm$0.01 & 0.57$\pm$0.01 &  1.71$^{+0.76}_{-1.28}$  \\
 kT$_{1}$ (keV)  & -- & -- &  0.94$\pm$0.02 & 0.93$\pm$0.02 &  \\
 kT$_{2}$ (keV) & 4.34$\pm$0.23 & 2.43$^{+0.76}_{-0.56}$ &  3.88$\pm$0.10 & 3.82$\pm$0.25 & 2.34$\pm$0.13 \\
 kT$_{3}$ (keV) & -- & 7.19$^{+6.14}_{-1.28}$ & -- & -- & 5.95$\pm$0.22 \\
 $\nu$          & -- & -- & -- & 1.2 & -- \\
 F$_{\rm tot} \times 10^{-10}$ erg cm$^{-2}$ s$^{-1}$
& 8.20 & 8.20 & 8.47 & 8.50 & -- \\
 F$_{0.2-10 keV} \times 10^{-10}$ erg cm$^{-2}$ s$^{-2}$
 & --  & -- & 7.67 & 7.61 & -- \\
 F$_{3-79 keV} \times 10^{-10}$ erg cm$^{-2}$ s$^{-2}$ 
& 6.08$\pm$0.60 & 6.13 & 6.25 & 6.79 & 6.35$\pm$0.27 \\ 
 F$_{\rm tot, unabs.} \times 10^{-9}$ erg cm$^{-2}$ s$^{-2}$
 & 0.96 & 1.31  & 2.38 & 2.47 & 1.30 \\ 
 F$_1 \times 10^{-10}$ erg cm$^{-2}$ s$^{-2}$
 & 2.94$^{+0.26}_{-0.13}$ & 3.31$\pm$2.18 & 7.74$\pm$0.54  & 0.77$\pm$0.08 & -- \\
 F$_2 \times 10^{-10}$ erg cm$^{-2}$ s$^{-2}$ 
& 6.08$\pm$0.06 & 5.95$\pm$0.12 & 7.69$\pm$0.27 & 7.44$\pm$0.86 & 3.53$\pm$0.04 \\
 F$_3 \times 10^{-10}$ erg cm$^{-2}$ s$^{-2}$ & --  & -- & 3.12 $\times 10^{-5}$ &  2.83$\pm$0.01 \\
 F$_{\rm pl}$ 10$^{-10}$ erg cm$^{-2}$ s$^{-2}$ & -- & -- & 0.29$\pm$0.14 & -- \\
Fe/Fe$_\odot$ & 1.35$\pm$0.66 & 0.47$^{+6.14}_{-1.28}$ & 0.33$\pm$0.16 & 0.35$\pm$0.01  \\
\hline
\end{tabular}
\end{center}
\caption{Parameters of the fits, with different models, to
 the spectrum of the overlapping  exposure time of NICER and NuSTAR,
 and to the averaged spectrum of all NuSTAR exposures: resulting
 value of $\chi^2$/(degrees of freedom), column density with and without covering fraction, plasma temperatures, index
 of the power law component, measured and unabsorbed flux, and iron abundance.} 
\label{tab:2}
\end{table}
\subsection{The upper limits for the non-thermal component}
 In all classical and recurrent novae previously  observed,  
 only  approximate estimates or upper limits were obtained for 
 the X-ray flux above 30 keV, always below 
 the expected upper limit for the non-thermal low energy ``tail''
  of the gamma rays \citep{Vurm2018}.
Therefore, the models addressed other manifestations of the shocks, 
namely flux and spectrum
 at radio wavelengths \citep[see][and references therein]{Vlasov2016}. Our {\sl NuSTAR} observation gives the opportunity to better verify and quantify the hard X-ray flux.

In all the fits in the first three columns of Table \ref{tab:2} and Figures \ref{fig_nustar_nicer_comp} and \ref{fig:nustar_ave}, there is a small excess at high energy. This excess disappears if we add a non-thermal power law component with a $\nu$=1.2 index (see the comparison in Fig. \ref{fig:non-thermal-fit}). However, both the $\nu$ slope and the flux normalization  of this component are unconstrained, and the addition is not necessary to improve the value of the reduced $\chi^2$  (shown as ``Model 2'' in Table \ref{tab:2}). Thus, the  resulting flux of the power law component, 2.9 $\times 10^{-11}$ erg cm$^{-2}$ s$^{-1}$ should rather be considered only as the upper limit for the flux of the possible non-thermal component.
\begin{figure}
\begin{center}
\includegraphics[width=140mm]{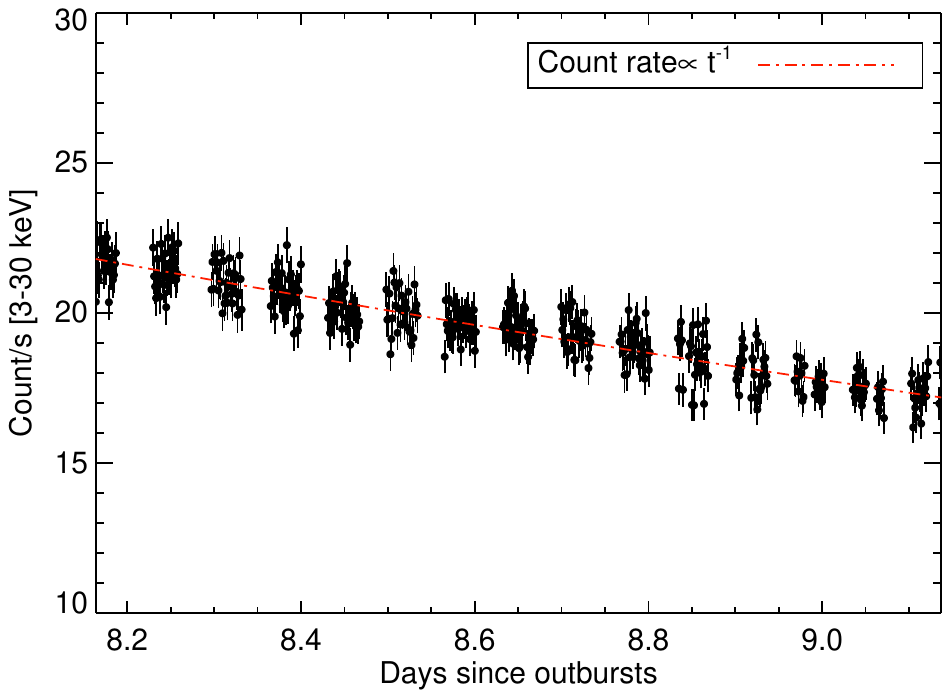}
\end{center}
\vskip -8cm.
\caption{{\em NuSTAR} FPMA light curve in the 3-30 keV energy range. The
 dot-dashed red line shows a power law fit: 
 the  count rate variation in time is proportional to $t^{-1}$.
 The FPMB lightcurve is exactly overlapping and we plot only one
 detector for clarity.}
\label{nustar_lc}
\end{figure}
\begin{figure}
\begin{center}
\includegraphics[width=130mm]{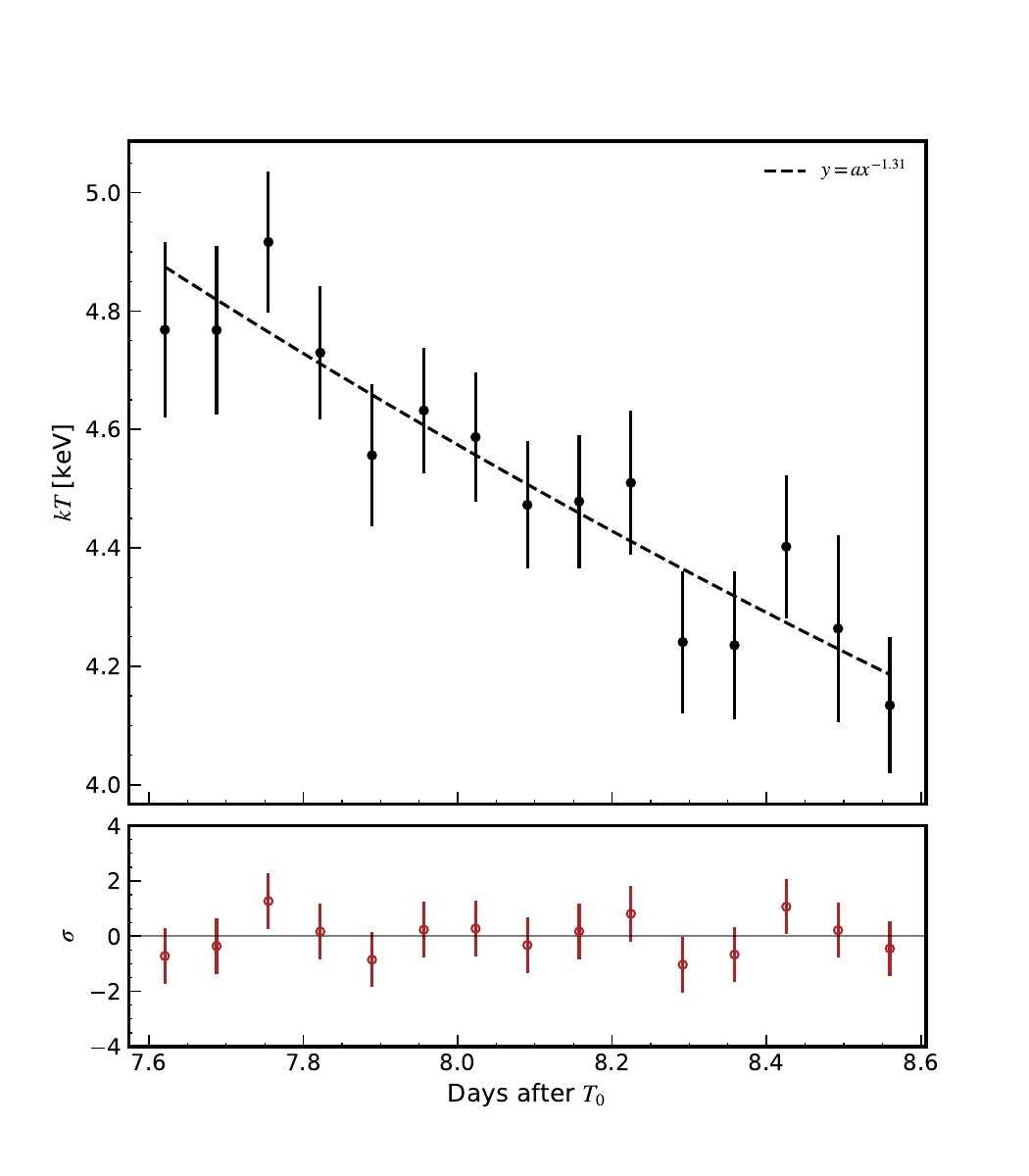}
\end{center}
\caption{Best fit temperature
 with one component APEC model, obtained
 for the single NuSTAR exposures between 2021 August 17 and 2021 August 18.
 The top panel shows how the trend can be fitted with t$^{-1.3}$.
 The residuals from this fit are shown in the bottom panel.}
\end{figure}
\begin{figure}
\begin{center}
\includegraphics[width=140mm]{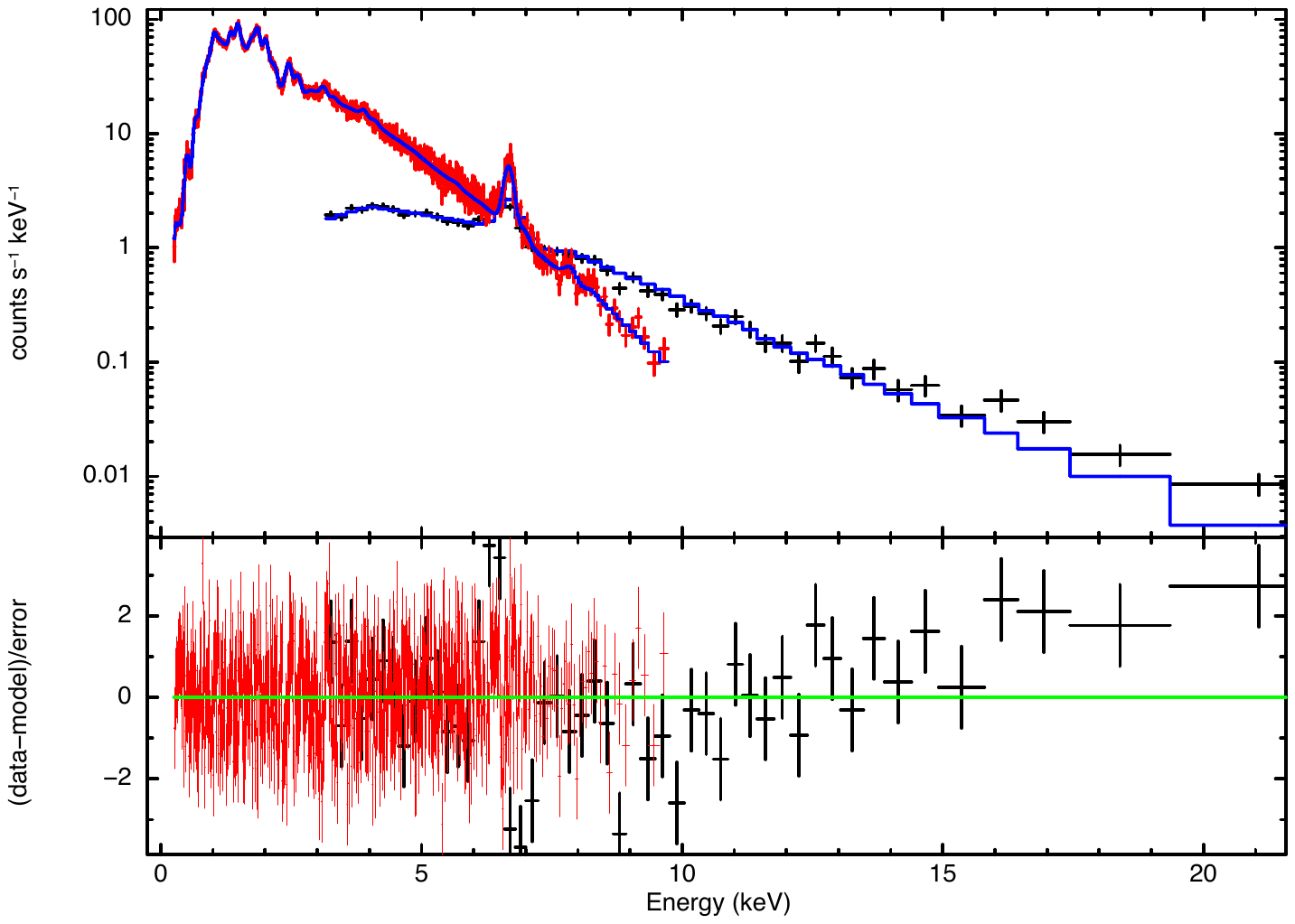}
\end{center}
\caption{The full-range spectrum observed on August
 17 2021 with NICER and NuSTAR in the overlapping time
 interval, in the 0.25-25 keV range
 and (in blue) the fit with the three component BVAPEC model,
  shown in Table 1. The residuals of the fit are shown in the lower panel.}
 \label{fig_nustar_nicer_comp}
\end{figure}
\begin{figure}
\begin{center}
\includegraphics[width=140mm]{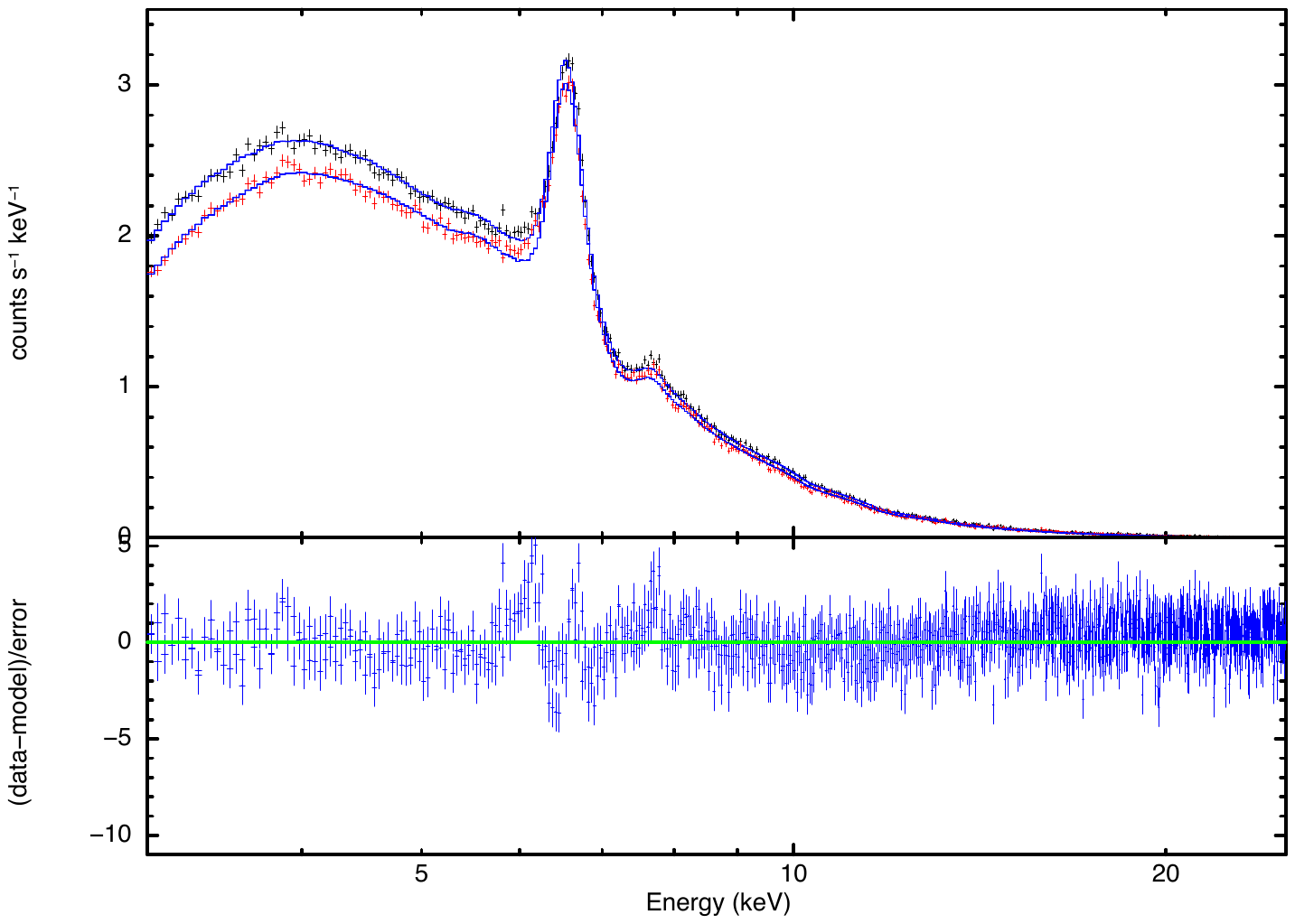}
\end{center}
\caption{The average NuSTAR FPMA (black) and
 FPMB (red) spectra avraged  
 over the whole time 1.2 days, and a fit with two vapec components at temperature
 of 2.65 keV and 11.70 keV, respectively, with
the parameters given in the first row of Table 1.}
\label{fig:nustar_ave}
\end{figure}
\begin{figure}
\begin{center}
\includegraphics[width=95mm]{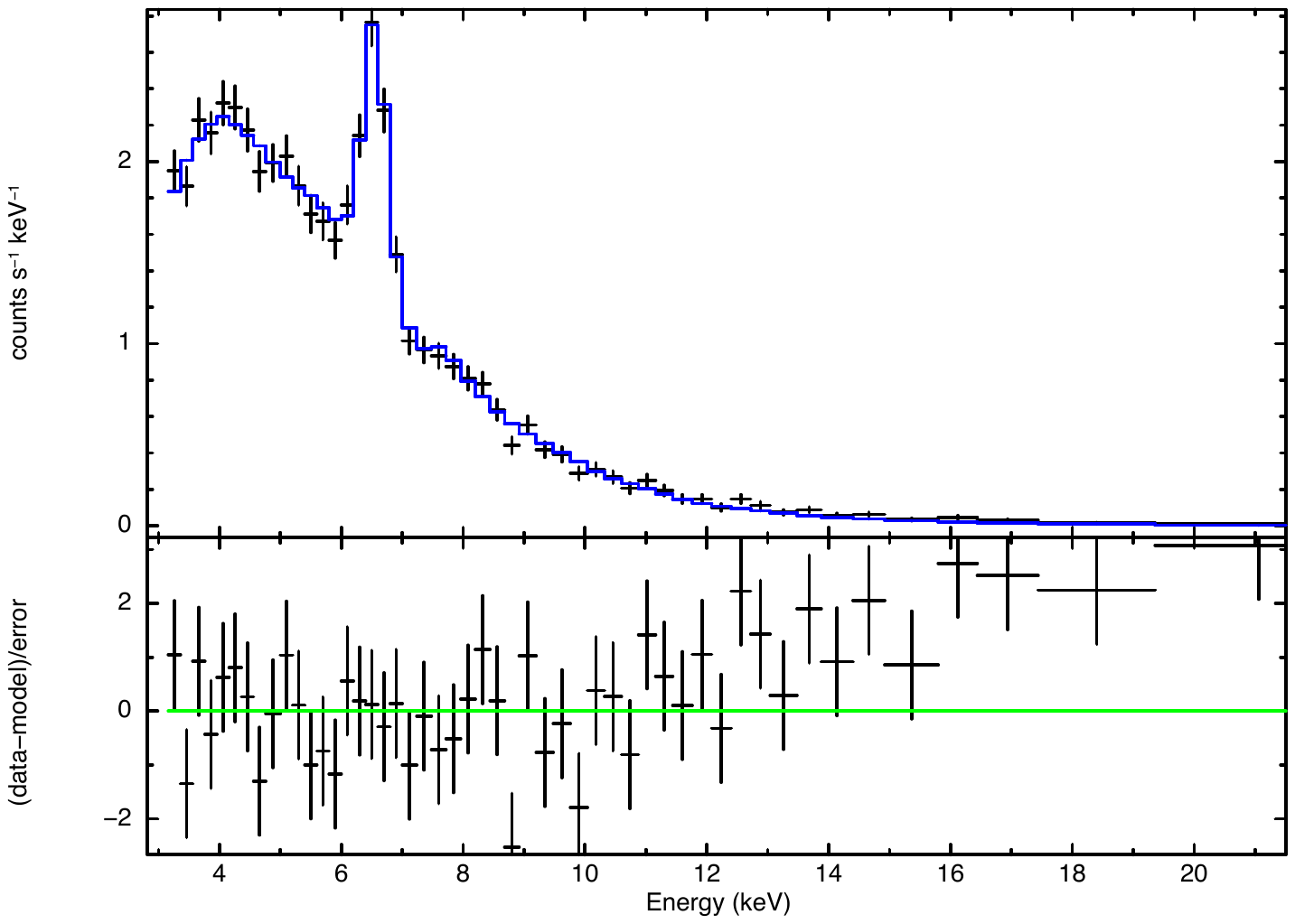}
\includegraphics[width=95mm]{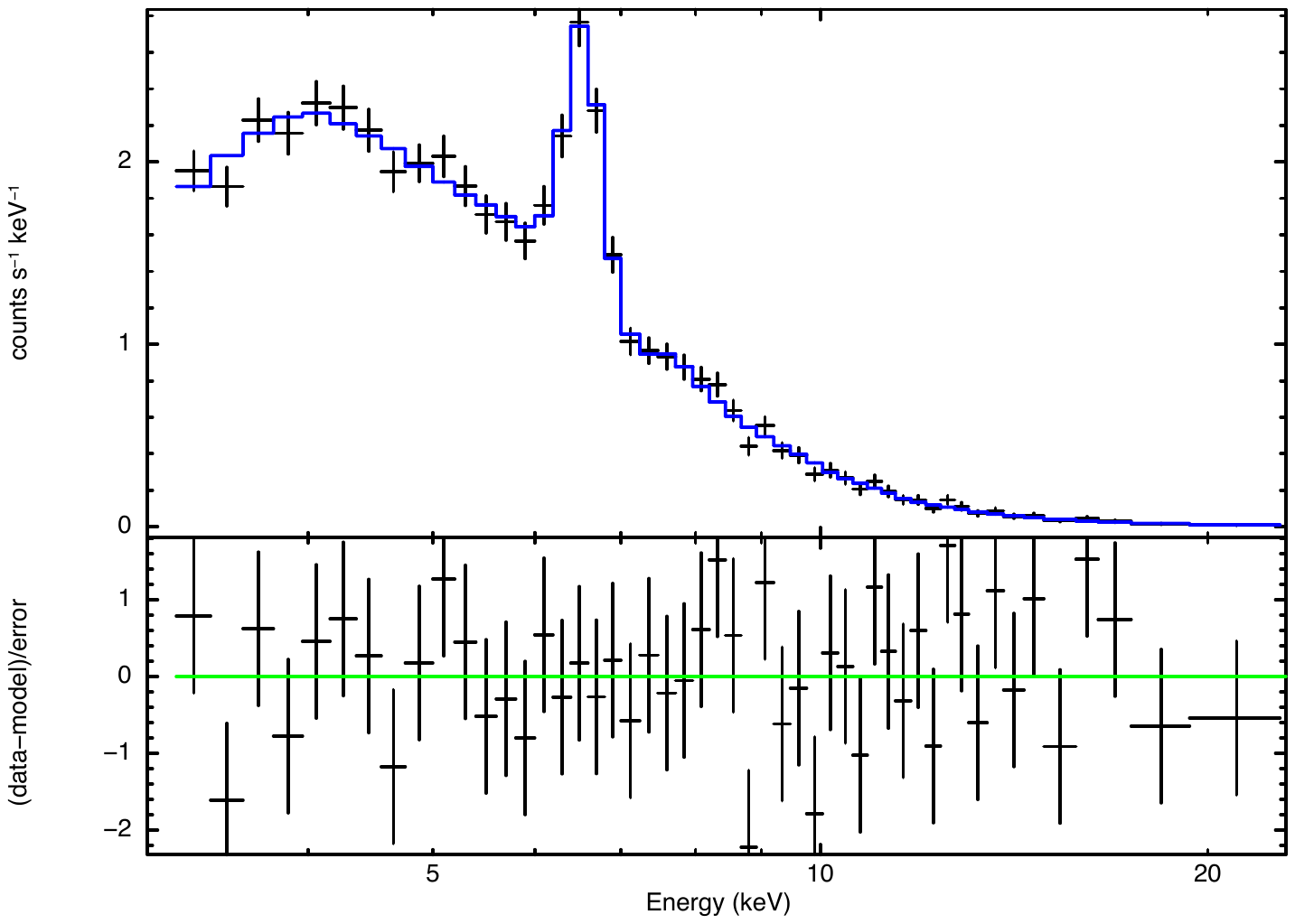}
\caption{The spectrum observed on August 17 2021 with NuSTAR FPMB in a time interval overlapping with  NICER, and the fit with the model shown in Fig. 4, with fixed N(H) of the partially covering absorber (see 2nd row in Table 1), with and without the power law component included in Model 2, last column of Table 1. Notice how the powerlaw improves the residuals at high energy.}
\end{center}
\label{fig:non-thermal-fit}
\end{figure}
By including the {\sl NICER} spectrum and thus fitting the whole energy range from 0.25 keV to 30 keV, we found out that a power law in addition to the thermal components
  cannot be ruled out. 
 However, the ``confidence contours''  (solutions within 3$\sigma$) 
 of flux normalization versus  power law are unbounded.

  We examined the possibility of a non-thermal component 
 considering a range that was  initially discarded in all fits
 because of extremely low signal-to-noise,
 namely  the 40-79 keV range. 
In the coadded exposures there is an almost 3 $\sigma$ detection in the FPMA, 0.00171$\pm$0.00043 cts s$^{-1}$, however the count rate is only 0.00069$\pm$0.00041 cts s$^{-1}$ in the FPMB. 
 Modelling the FPMA spectrum
 {\it including the power law component} we found that the thermal,
 {\sl APEC} flux would contribute close to 75\% of the flux even at energy higher than 40 keV. 
 In practice, this translates in an upper limit to the non-thermal flux
  in the range  40-78 keV
 of 10$^{-12}$ erg cm$^{-2}$ s$^{-1}$, which,  like for other
 novae observed with {\sl NuSTAR},
 this is not a significantly constraining value. The gamma-ray
 flux measured with Fermi on the same day was 
 F$_{\rm \gamma} \simeq 7 \times 10^{-10}$ erg cm$^{-2}$ s$^{-1}$.
\section{Was the initial X-ray luminosity absorbed or suppressed?} \label{sec:disc}
The flux in the {\sl Fermi} energy range peaked about 1 day post-optical maximum, with a value
 of F$_{\rm GeV \gamma} \simeq 5 \times 10^{-9}$ erg s$^{-1}$ cm$^{-2}$, while the flux measured with {\sl H.E.S.S.} peaked on days 4-5 \citep{HESS2022}, at
 F$_{\rm TeV \gamma}\simeq 8 \times 10^{-12}$ erg s$^{-1}$ cm$^{-2}$. 
Only the {\sl H.E.S.S.} peak flux and epoch of maximum are
 compatible with the X-rays, 
 because only a small fraction of the shock power accelerates particles,
 less than 1\% \citep[see][]{Martin2013}. Thus,
 we suggest that the coincidence of the shock luminosity measured
 in X-rays and the effects of particle acceleration 
 measured with {\sl H.E.S.S.} have a physical meaning and point at the
 same event. 

We need to consider that we only have reliable values for the measurable absorbed X-ray flux, while the parameter of interest for the comparison, namely the value of the {\it unabsorbed} X-ray flux, is model-dependent. However, even the 
highest estimate of the X-ray maximum unabsorbed flux, 
obtained with the model of \citet{Orio2023}, F$_x \simeq 2 \times 10^{-8}$ erg s$^{-1}$ cm$^{-2}$, 
 is a factor of 5000 larger than the H.E.S.S. flux, but  only a factor of 2
 larger of the Fermi flux.  With the lower values obtained assuming the models of \citet{Page2022} and \citet{Islam2024}, the unabsorbed
 X-ray flux at the source would still be two orders of magnitude greater than measured with {\sl H.E.S.S.}.
 We mentioned the phenomenon
 of differential $\gamma$-ray absorption of \citet{Phan2025};
 we see a problem in that it  does not take the X-ray phenomenology into
 account. The two-shocks model proposed by \citet{Diesing2023}
 is instead compatible with two possible cases: 
 
 $\bullet$ An initial, strong shock - in a different site than the shock we
  mainly observed in X-rays -  was 
 absorbed by extremely high column density {\it ahead of it
 and along the line of sight}, even if the density 
  per volume unit where the shock occurred was not
 high enough to prevent particle acceleration.
 If this was the case, there may have been at most a very small contribution 
 to the hard X-ray flux, and none at all at energy below 30-40 keV.
 
  $\bullet$ The second possibility is the model by \citet{Metzger2025}: 
 turbulence mixed the hot plasma with much cooler gas, 
balancing the shock heating and greatly reducing the volume of hot gas
 that emitted X-rays. In this case, there must have
 been more than only one shock, but the X-ray flux associated with the
 ``{\sl Fermi} event'' may have partly contributed to the total X-ray flux from
 the beginning, since the X-ray spectrum would not be changed while
 the flux was reduced.
  
Simulations of the spectrum with the PIMMS tool of HEASOFT indicate that
  an X-ray flux at least 100 times higher than the gamma-ray flux 
measured with {\sl Fermi} would have been in very large part absorbed only by a column density N(H)$\geq 10^{24}$ cm$^{-2}$,
 an order of magnitude larger than the upper
 limit to the column density estimated with any of  the models 
 used to fit the  X-ray spectra of the first 5 days.
  If X-ray absorption was the reason
 for the low X-ray flux, the X-rays we measured were always associated 
 with the ``{\sl Cherenkov} event'' and not
 with the ``{\sl Fermi} event''. 
On day 9, a flux of  $\approx 5 \times 10^{-9}$ erg s$^{-1}$ cm$^{-2}$ 
 was still measured in the Fermi range, almost an order of magnitude
larger than the flux in the whole 0.2-79 keV range {\sl NuSTAR+NICER}.
The {\sl NuSTAR} upper limit to the flux above 40 keV is compatible 
 with an unabsorbed flux at least 100 times larger than the {\sl Fermi} flux only for a column density N(H) $> \times 10^{25}$ cm$^{-2}$, clearly ruled out 
 because a soft X-ray component was measured with {\sl Swift} and with {\sl NICER}..
 
Thus, in this case the ``{\sl Fermi} event'' must have had occurred where the 
 the particle density was so high as to inhibit particle acceleration at TeV
 energies but not at GeV  energies.
The relevant timescales to examine are the one for proton-proton loss, versus the timescale for particle acceleration. The latter must be shorter than the first, or else the particle acceleration is suppressed. 
Assuming Bohm diffusion ($D_B = cr_L/3$, where $r_{L}$ is the particle gyroradius), and that the acceleration time is related to the diffusion time as $t_{acc} \simeq 8t_{diff} = 8D_B/v_s^2$ \citep[e.g.,][assuming a compression ratio of 4]{Drury1983}, we obtain an acceleration time, $$ t \approx 24 \text{s} \times \bigg(\frac{E}{\rm GeV}\bigg)\bigg(\frac{B}{\rm G}\bigg)^{-1}\bigg(\frac{v_s}{1000\text{ km s}^{-1}}\bigg)^{-2}$$.

From equation 3 of \citet{Diesing2023} the proton-proton loss time is $$ t_{\rm pp} \approx 260 \ {\rm days} \times \bigg(\frac{ n_{\rm 0}}{10^8 \rm cm^{-3}}\bigg)^{-1} $$ where n$_{\rm 0}$ is the number density of the ambient medium in front of the shock.
   
With electron density $n_e\leq 10^8$ cm$^{-3}$ 
 the proton-proton loss time is $\geq$260 days,
 while assuming a 1 Gauss magnetic field and shock velocity
  of 4500 km s$^{-1}$
 the acceleration timescale for GeV energy is only 1.2 s. 
However,  the acceleration timescale is 1000 times longer for energy of 1 TeV; 
 it becomes very difficult to accelerate particles to TeV energy with 
 density $n_e > 5.3 \times 10^{8}$ cm$^{-3}$, as the proton-proton loss time becomes comparable to the acceleration time, while with
 the same density particles are still
 accelerated at GeV energy (\citet{Diesing2023}
 assumed a shock velocity of only 1300 km s$^{-1}$ considering that
shocks in the impact with a dense medium expand more slowly 
 on the equatorial plane and obtained n$_e \geq 2 \times 10^{10}$ cm$^{-2}$,
 but  
 this may hardly be the case for novae in symbiotics
 \citep[see][]{Shen2022},
  so our lower value of n$_e$ may be more realistic).
In any case \citet{Orlando2009} proposed that the
 ``density enhancement'' in the equatorial plane,
 even quite far from the red giant, may have been anywhere
 in the density range of few 10$^{7}$  cm$^{-2}$ to few  10$^{10}$cm$^{-3}$, 
 so it is indeed conceivable
 that acceleration to TeV energy was impossible where the initial shock
 happened.   

 Assuming that the proton/ion particle density is the same as
 the electron density,  there must have been
 a much denser medium ahead of the shock and along
 the line of sight to yield sufficient column density to
 absorb practically all the X-ray flux associated with 
an initial, very powerful shock whose consequences were 
  observed with {\sl Fermi} {\sl NOT} to be observable in X-rays. 
The red giant subtends a large angle and was reached
 by the outflow after about half a day. From \citet{Ludwig2012},  we learn that the particle density at the surface of a red giant is of the order of 10$^{16}$ cm$^{-3}$,  and is still of a few 10$^{14}$ cm$^{-3}$ at a distance of 7 $\times 10^{11}$ cm the photosphere, thus a column density above 10$^{25}$ cm$^{-2}$ 
 can be present ahead of the shock  if it occurred close to
 the red giant atmosphere.  A red giant does not have
 a sharp transition to a dense surface; 
 there is instead an extended zone with a very sharp gradient
 in density, a layer around
 the star,  in which X-rays  may have been in large part absorbed by
 the dense wind. On the other hand, in the 2021 outburst,
 the red giant was not completely along the line of
 sight: with respect to us it was at orbital
 phase 0.73, near the 2nd quadrature and approaching us,
 so it may have only partially shielded the X-rays.
 However,  most likely there was also  an additional
 dense zone that was not included, for
 instance, in the model of \citet{Orlando2009}: 
  simulations by \citet{Booth2016} of the RS Oph binary show 
 a dense trail in the equatorial plane,
 which at the time of the 2021 outburst, at the 2nd quadrature,
 would have been facing line of sight. The inclination of RS Oph is 
 about 50$^o$ \citep{Brandi2009}, so this contributes
 significantly to the absorption. Quantifying the exact
 column density at each orbital phase is difficult, but we 
note that \citet{Dumm2000} modeled UV observations
 of the wind accretion wake of the detached eclisping binary and
 symbiotic RW Hya
 with a column density along the line of sight {\it inside the binary}
 of almost 10$^{24}$ cm$^{-2}$ at a given orbital phase 0.7.  
  
   The second  case, that of turbulence and impact with a cold medium, 
 is even more likely to occur only in  the vicinity of the red giant, 
 where  the outflow
    would encounter a thin shell of dense, cool wind, 
 not exceeding  the photospheric temperature of
    4000 K (corresponding to kT$\simeq$0.35 eV) \citep{Ludwig2012}.
    So, there was a very large temperature  difference,
 and most likely also a sharp density contrast between the outflow
 and the wind layer very close to the red giant.
      This is similar to the thin, cold and dense shell in which
      the slower outflow in a classical, short period nova may be impacted
       by a faster outflow even
 a few weeks after the optical maximum in the model of \citet{Metzger2025}.
        In   the case of novae occurring
 in symbiotics, the situation is different
 because the shocks occur almost immediately, 
 since in these novae the radiation pressure driven wind is the
 first and probably only outflow mechanism \citep{Shen2022}. 

 If the X-ray flux in the initial shock was {\it suppressed}, 
 and not {\it absorbed}, the spectrum may have been unvaried 
 and the reduced flux from this shock contributed to the soft X-ray 
 flux below 10 keV in the first 4 days. 
 In the {\it X-ray absorption case}, 
 in 2006,   near the 1st quadrature the trail was behind
 the red giant, but we did not have measurements in the Cherenkov range
 to assess the possible second shock.
 In the next outburst of RS Oph, depending on the orbital
 phase at which it occurs and making a comparison
 with the X-ray and gamma-ray lighcurves of 2021, we should be able to 
 finally understand whether the {\it absorption} or the {\it turbulence}
  case 
 explains the ``mystery'' of the low X-ray flux compared to the flux
 of the ``{\sl Fermi} event''.
\begin{longrotatetable}
 \begin{deluxetable*}{lrrrrrrrrrrrrrl}
\tablecaption{Physical parameters of the five best known Galactic
 novae in symbiotics , including components' masses,
 secondary spectral type, recurrence time, recorded 
 maximum outflow velocity in outburst, time for a decline of 2 and 3 optical 
magnitudes from maximum, time (in days) for the onset
 of the SSS flux and time for its decline by $\simeq$70\%, measured
 or estimated peak X-ray flux (0.3-10 keV) and peak gamma-ray flux
 (in the Fermi energy range, also few TeV range for RS Oph), and
 estimates of the unabsorbed X-ray flux in the 0.3-10 keV range. 
}
\tablewidth{1100pt}
\tabletypesize{\scriptsize}
\tablehead{
\colhead{Name} & \colhead{m(WD)} & \colhead{m(giant)} &
\colhead{spectral} & \colhead{d} & \colhead{P$_{\rm orb}$} & \colhead{t(rec)} &  \colhead{V$_{max}$} & \colhead{v(ej) } & \colhead{t$_2$}  & \colhead{t$_3$}  &  \colhead{times(SSS)} & \colhead{F$_x \times 10^{-10}$}  & \colhead{F$_\gamma \times 10^{-10}$}  & \colhead{F$_{\rm x,u} \times 10^{-10}$} \\
\colhead{}     & \colhead{M$_\odot$} & \colhead{M$_\odot$}   &
\colhead{type} & \colhead{kpc}   & \colhead{days}          & \colhead{years}  & \colhead{} & \colhead{km s$^{-1}$} & \colhead{days} & \colhead{days} & \colhead{days} & \colhead{erg/cm$^{2}$/s} & \colhead{erg/cm$^{2}$/s} 
 & \colhead {erg/cm$^{2}$/s}}
\startdata
 RS Oph     & 1.2-1.4 (1)  & 0.68-0.80 (2) & M0-2 III (2) & 2.40$_{-0.28}^{+0.12}$ & 453.6 (2) & 9-21 & 4.8 & 7550 (3) & 7 & 14 & 32/52 (4) & 13.60 (5) & $\approx$40 (GeV) (6)  & 220 \\
  &  & & & & & & & & & & & & $\approx$0.08 (TeV) & \\      
 V3890 Sgr  &  1.35$\pm$0.13  &  1.05$\pm$0.11 (7)    & M5 III (7)  &
  $\approx$ 9 (7) 
 & 747.6 (7) & 28     & 7 & $\geq$ 4200 (9)  & 6 & 14 & 8+16 (10) & 0.77-0.97 (10) &  $\approx$2 (11) & $\approx$2-2.5 (10) \\
 & (7,8) & & & & & & & & & & & & & \\
 V745 Sco   & 1.4 (12) &          & M4$\pm$2 III (13) & 5.88$_{_1.38}^{+1.92}$ & -- & $\approx$25 & $\approx$8.5  & 4250 (14) & 2 & 5  & 4+6 (15) & 1.20  & $\leq$4 (17) &  $\approx$1.9 \\
 T CrB      & 1.37$\pm$0.01  & 0.69$\pm$0.02 (18) & M4 III (18) & 0.887$^{+18}_{-29}$ & 227.55 (18) & 80  & 2 &  & 4 & 6 & -- & -- & -- & -- \\
 & (18) & & & & & & & & & & & & & \\
 V407 Cyg   &   -- & -- & Mira (19) & 2.87$^{+0.60}_{-0.69}$ & 43$\pm$5 yrs & --& 8.31 & 3500 (20) & $\simeq$32 & $\simeq$63 & $<4$? &  0.12 (21) & $\approx$5.6 
(22) & $\approx$4 \\
 &  & & & (19) & (19) & & & & & & & & & \\
\enddata
\tablecomments{ (1) Nelson et al. 2008; (2) Brandi et al. (2009); (3) Munari et al. (2022) - note that all following measurements are at or below 4500 km/s; (4) Osborne et al. (2011); (5) Orio et al. 2023; (6) Cheung et al. (2022) and HESS Coll. (2022)
  (7) Mikolajewska et al.  (2021); (8) Page et al. 2020;  
 (9) Munari \& Walter (2019)
 and Strader (2019); (10) Orio et al. (2020), Page et al. 2020;  (11) Buson et al. (2019)
 (12) Shara et al. (2018);
(13) Harrison et al. (1993); (14) Anupama et al. (2014) ; (15) Page et al. (2020);
 (16) Orio et al. (2015); (17) Cheung et al. (2014); (18) Hinkle et al. (2025);
 (19) Munari et al. (1990); (20) Giroletti et al. (2020); (21) Nelson et a. (2012); (22) Abdo et al. (2010). The V745 X-ray fluxes were estimated by us from Swift observations. The peak unabsorbed X-ray flux of V407 Cyg was inferred
 from the emission measure given by (21). }
\end{deluxetable*}
\end{longrotatetable}
\section{Comparison with other novae known to be in symbiotics} 
 The first  nova in a symbiotic
 that was well monitored in both gamma-and X-rays was 
  V407 Cyg. The known parameters of this nova are in the last row of Table 1.
 The distance is still uncertain, but it is likely close to 
 that to RS Oph: although the GAIA DR3 geometric distance is
  4.63$_{-1.33}^{+1.85}$ kpc, the photogeometric
 distance of 2.87$^{+0.60}_{-0.69}$ kpc \citep{Bailer2021} 
 is in the range of 2.5-3 kpc suggested by
 by \citet{Munari1990} using infrared data.  \citet{Nelson2012} adopted
a value of  2.9 kpc.

Like in RS Oph, the gamma-rays peaked at the beginning, after about a day
 and almost coincidentally with the optical maximum, but the
 shock in this case has been attributed
 to the impact with the wind of the Mira companion \citep{Munari1990}, not with
 the giant 
 itself, which is at a large distance, about 25 AU.
 A Mira can loose up to 10$^{-4}$ M$_\odot$ yr$^{-1}$ \citep[see][]{Bowen1991}
 and its surrounding wind is expected to be much denser than that of red giants
 of M III spectral type, like in RS Oph.  
The gamma-ray flux that was about a third
 than in RS~Oph case \citep{Fermi2010, Martin2013} and fell below detection
 threshold after 2 weeks.  {\sl VERITAS} observations detected no flux in the TeV range, with an upper limit of 2.3 $\times 10^{-12}$ erg cm$^{-2}$ s$^{-1}$ \citep{VERITAS2012}, several times lower than the RS~Oph peak measured with {\sl H.E.S.S.} and {\sl MAGIC}; however, this measurement  was obtained only between days 9 and 12 of the outburst.
 For V407 Cyg it is instead the rise of the X-ray flux to its peak value 
that has been  attributed to the impact with the Mira atmosphere. It 
was always about two orders of magnitude less than the RS~Oph 
 measurement \citep{Nelson2012}, but it 
 peaked suddenly after about 3 weeks, which - given the
 expansion velocity of the ejecta - is about the time for the outfow
 to reach the Mira. The peak was followed by a gradual decay.

 An important difference between V407 Cyg and RS Oph and the other
 {\it recurrent} novae  known in symbiotic
 systems (the first four entries in Table 3)
 is that in V407 Cyg, given
 the orbital separation, the presence of  an accretion disk is ruled out.
 Another striking difference is the absence of a luminous SSS phase.
 A break in the X-ray 
 lightcurve was interpreted by \citet{Nelson2012} as due to an absorbed 
 low luminosity soft component turning off, however repeating the modeling
 we find that a much better fit is obtained with 
 a thermal APEC component in the 50-90 eV range  in different dates;
 we do know that in novae 
 shocks can generate also a very soft thermal
 spectral component \citep[e.g.][]{Mitrani2024, Mitrani2025}. 
It is thus conceivable that the SSS phase of this nova was
 really extremely short, shorter than the longest interval of 4 days
 lapsed between {\sl Swift} pointings. 
The  duration of the SSS is often interpreted as commensurable to the
 duration of the nova wind, but in this case,
  it is in stark contrast with the long lasting optical
 brightness indicated by t$_2$ and t$_3$. 
 A possible explanation is that the accumulated envelope
 before the outburst was much smaller than in RS Oph, while the shock- or photo-ionized
 ejecta were optically luminous for a long time.

\citet{Nelson2012} found that 
before the forward shock expanded and cooled, the peak post-shock temperature
 was  higher than 12 keV; at the time
 the wind probably reached the Mira, the highest shock temperature 
 could not be well determined, except  for being
 in a wide range between 2 and 6 keV. 
\citet{Nelson2012} also modeled the X-ray lightcurve of
 V407 Cyg with a a circumstellar material wind distribution as $r^{-2}$, where $r$
 is the distance from the Mira, and two shocked plasma components at different temperature, necessary to explain the X-ray spectra.
Their model shows that
the peak of the X-rays occurred when the Mira was reached by the outflow because the wind material became
 much denser. Of course, one must keep in mind that the Mira subtends
 a small angle in comparison with the red giant of RS Oph, explaining the much
 lower peak of the X-ray flux when the giant was embedded in the nova
 outflow. \citet{Nelson2012} 
 suggested that the density near the Mira must have been so high to inhibit particle acceleration, so that the gamma-ray ceased as the ejecta reached the companion
 (this is exactly the opposite of what we proposed for RS Oph).
\citet{Orlando2012} computed a full-fledged hydrodynamic simulation
 with the wind density distribution proposed by \citet{Nelson2012}, a density 
enhancement in the equatorial plane, and the outflow collimated 
 in the polar direction.  
 However, neither work addressed the different magnitude of 
the gamma- and X-ray flux
 at the beginning, which remained an open
 problem: the only solution seems to be
 that turbulence -as described by \citet{Metzger2025} ensued in the impact with
 the cool and slow Mira wind.  One issue of these models is
 that they may have assumed too low values of density in the equatorial
 plane and mass loss rate from the Mira: only 10$^{-7}$ M$_\odot$ yr$^{-1}$,
 however a Mira may have a much higher mass loss rate
  \citep[e.g.][]{Bowen1991}.

From the  comparison of RS Oph
 with V407 Cyg we understand  that the properties
 of the outburst are greatly shaped by the environment of the
 symbiotic and the binary parameters. It is thus
 interesting to extend the comparison to the other systems in Table 3 that are
 more similar to RS Oph, V3890 Sgr and V745 Sco, even if the X-ray and gamma-ray
 monitoring was sparser.  The orbital period of
 V3890 Sgr is of 747.6 days; it is not known for V745 Sco.
  In both novae the spectral  class of
 the red giant is a later type than for RS Oph, so they may  have a larger
 mass loss rate in the wind because they are more
 luminous \citep[e.g.][]{Sanner1975}. The distance 
is possibly about triple than to RS Oph for
 both these novae, albeit with large uncertainty: Table 3 reports for
 V3890 Sgr the approximate distance obtained assuming Roche-Lobe
 filling \citep{Mikolajewska2021}, however this is uncertain
 and the GAIA distance after the DR3 is 4.36$_{-2.64}^{+1.31}$ kpc.

 Despite substantial difference in duration
 of optical and SSS luminosity, both much shorter than in RS Oph,
 the X-ray flux
 attributed to shocks and the gamma-ray flux in V3890
 Sgr followed a similar trend to RS Oph: gamma-ray peak within a day,
 and X-ray flux peaking after almost 5 days since the optical maximum.
 Given comparable
 ejection velocities, in V3890 Sgr the outflow should have reached the
 red giant in about one day. 
 The lower values of fluxes in gamma- and X-rays seem to be compatible
 with the  
 smaller angle subtended by the red giant for the gamma-rays, and
 to a more diluted wind at a larger distance from the giant for the
 X-rays. Moreover, the inclination of the orbital plane
 is higher, about 70$^o$ for V3890 Sgr and 60$^o$ for V745 Sco,
 compared with about 50$^o$ for RS Oph.  
 Another interesting aspect to consider is the configuration at the time
 of the outburst: in V3890 Sgr almost at first quadrature, with the trail 
 behind the line of sight, in V745 Sco, with both the red giant and
 the trail along the line of sight.
 
 A brief and early
 SSS phase for  both these two novae prevented disentangling the X-rays' spectrum
 due to the shocks clearly enough to obtain a good spectral fit
 after day 9, but from Fig 5 of \citet{Ness2022} we 
 infer that the evolution was similar to RS Oph, with a hard-ish 
 component of the X-ray flux cooling slowly and lasting for a few weeks.
 The same can be said for V745 Sco, which however plateaud in X-rays
 after a day; the SSS rise occurred after 4 days, with a SSS maximum 
lasting for only 2 days and slowly decaying for the following 8 days,
 after which a hot thermal component had a temperature of
 1.2 keV the {\sl Swift} archival spectra. 

 \citet{Shara2018} computed
  models that explain several properties of these two novae and of RS Oph
 with a small but significant difference of WD mass in an extreme range:  
 1.38 and 1.40 M$_\odot$ respectively for V3890 Sgr and V745 Sco, versus
 a value of 1.31 M$_\odot$ necessary to explain the RS Oph
 outburst. This small difference at the extreme end
 of the distribution of WD masses is sufficient to justify 
 the shorter periods of luminosity in optical and especially
 in SSS X-rays, while the longer recurrence times
 are modeled with a slower mass accretion rate onto the WD 
 (probably due to the larger orbital separation than in RS Oph).
 In the set of models by \citet{Shara2018}, like in other similar work,
 a nova event with a WD mass $<$1.1 M$_\odot$ ejects even a little
 more than the mass it accretes, but a massive 
 WDs mass may eject less than the accreted mass and be increasing in mass,
 depending on the other physical parameters.
 Specifically, RS Oph is modeled as having accreted
 only a mass of  $10^{-6}$ 
 M$_\odot$ while it has accreted an envelope of 1.65 $\times 10^{-6}$ 
 for the outburst to be triggered. Since it does
 not eject all its accreted envelope, the following
 outbursts may be spaced closer apart. The discrepancy
 between accreted and ejected mass
 is even larger for V3890 Sgr ad V745 Sco,
 that are supposed to be already close to a final event:
 either a type Ia SN explosion or an accretion induced collapse.

\subsection{The predictions for a new T CrB outburst}
The models by \citet{Shara2018} indicate that
  T CrB would accrete 1.67
 $\times 10^{-6}$ in 80 years  and eject even more mass than accreted,
 so it may not be on the same path to a type Ia SN as the other three
 well studied recurrent novae in symbiotics in the Galaxy, namely
 RS Oph, V3890 Sgr and V745 Sco.

 T CrB had recorded outburst in 1866 and at the end of 1945, with an elapsed time of 
 almost 80 years between them. In 2026, 80 years
 will have elapsed again.
 In addition, a brief luminosity dip that the nova experienced, similar
 to one observed before the 1945 event, followed by
 increase brightness and signs of activity both in optical
 and in X-rays, seemed to indicate that the new eruption may really
 be close. This has prompted a flurry of reserach activities focused
 on better determining the parameters of the system, including 
 the mass accretion rate and its irregularity \citep[e.g.][]{Luna2019,
 Hinkle2025}, and on calculating
 dedicated models \citep{Jose2025, Starrfield2025, Wallace2025}. A super-remnant
 has also been discovered, the relict of past outbursts, extending
 out to 30 pc in diameter \citep{Shara2024}.
 
  Because of the closer distance to us ($\leq$900 pc, see Table 3) the new
 outburst is expected to be spectacular, at least at optical wavelengths
 because the previous outbursts reached V$\simeq$2, so the nova will
 be visible with the naked eye. The detection at high energy of
 RS Oph with the Cherenkov
 telescopes has spurred interest in possible detection of 
 neutrino flux. \citet{Thwaites2025} recently wrote: {\it ``Due to its closer distance
  and higher optical flux, which has been well measured in two historical eruptions, the expected neutrino signal from T CrB is several times stronger than that
  from RS Ophiuchi. Furthermore, T CrB is located in the Northern sky at
 a declination where IceCube's sensitivity is an additional factor of a few
  better than at the location of RS Ophiuchi, which is beneficial
 to this search''}.   

 Members of our group have simulated high-resolution X-ray spectra
 assuming an initial X-ray luminosity due to shocks 10 times larger than  
 in RS Oph, about the same temperature, and contemplating the
 possibility of a faster cooling, at a rate possibly 3 times that of
 RS Oph (since the whole nova evolution seems to be faster and the outflow
 may end sooner). Assuming the same spectral
 characteristics and  a similar rise time as in
 RS Oph, during the first 4 days that {\sl XRISM} spectra 
 will have unprecedented   
 resolution in the region of the iron and silicon lines between 6.4 and 8 keV,
 such that the ionization time scale would be well determined and
 the peak temperature and unabsorbed flux will be estabilished with little
 uncertainty (poster paper presented by us  
 the ``XRISM 2025 International Conference'' in
 Kyoto, Japan in 2025 October).
  However,  \citet{Orlando2025} recently  
 made quite different predictions. Their simulations of the {\sl XRISM} and
 {\sl XMM-Newton} RGS spectra indicates that the X-ray flux at energy above 
 2 keV would not be larger than in the case of RS Oph, despite
 the closer distance, and that by day 6 
 it would barely be detectable with {\sl XRISM}. 
  In the softer RGS range, these authors 
 predict that the X-rays would be observable 
 for up to about a week. Their Fig. 7
 shows a spectrum that is less luminous than that of RS Oph,
 with much broader emission lines. 
  
  These authors  assumed 
  an extremely different binary  environment that in RS Oph, 
  based on  radio observations done
 by \citet{Linford2019}, in which the 
 detected radio flux was quite lower than in 
 RS Oph and other symbiotics. \citet{Linford2019}
 in fact found that, even if the radio emission had increased since 
 observations done a few years earlier, the spectral energy 
distributions were consistent with optically thick thermal
 bremsstrahlung emission from a photoionized source. \citet{Orlando2025}
 derived the wind density from these data,   
 assuming  that the wind was fully ionized, and concluded 
 that the {\it retained} mass loss rate of the giant (the net portion
 of the wind mass loss rate that remains in the CSM) in 
T CrB  is of only
 about 4 $\times 10^{-9}$ M$_\odot$ yr $^{-1}$. Although
 these authors
  considered also the possibility of
 a  somewhat higher mass loss rate and  the presence
 of a density enhancement in the equatorial plane and of an accretion disk
 (the early phase would be dominated by shocked disk material), their
 3-D simulation of the circumstellar wind shocked by the nova resulted 
 in a less violent
 shock and a softer X-ray spectrum than RS Oph, with much less
 flux at high energy. For comparison, from the ionization
 structure of the CSM around RS Oph, \citet{Booth2016} estimated
 a red giant mass loss rate of 5 $\times 10^{-7}$ M$_\odot$ yr $^{-1}$.
 In this  model, T CrB would probably not be a 
 detectable gamma-ray source
  at all and this would also bear importance on
 the possible measurement of a flux of neutrinos; while
  instead if shocks similar to those in RS Oph  occur
 at T CrB distance
 there is a significant probability of an IceCube detection
 \citep{Thwaites2025}, the parameters adopted in the \citet{Orlando2025}
 would rule it out. 

 Radio observations were repeated by \citet{Petry2025}
 with {\it ALMA} and these authors found that in 2024 the wind was far from being
 fully ionized, but the radio flux significantly decreased and 
 they estimated a lower radio spectral index than in 
 the measurements of \citet{Linford2019}. \citet{Petry2025}
 substantially confirm also the radio flux measurements of 2016 and favor
 a red giant mass loss rate of about 10$^{-8}$ M$_\odot$ yr$^{-1}$.  
 However, we note that the large number of optical spectra available in
 the ARAS database (see \url{https://aras-database.github.io/database/conditions.html} and  
 \url{https://aras-database.github.io/database/tcrb.html}), and 
 also spectra published by \citet{Munari2016, Maslenni2023, Stoyanov2025}  
 show the complete absence of forbidden nebular lines for T CrB 
 at quiescence, 
  like in RS Oph,  a fact that translates in lower limits of
 n$_e > 10^5$ cm$^{-3}$ for most lines in the optical range, and as high as 
 n$_e > 10^{7.5}$ cm$^{-3}$ for [O III] $\lambda\lambda$4363.2
 \citep{Appenzeller1988},
 in stark contrast with the n$_e > 10^3$ cm$^{-3}$
 wind density assumed by \citet{Orlando2025} and even with their enhanced equatorial density of n$_e > 10^6$ cm$^{-3}$.
 Furthermore, in \citet{Luna2018} the mass accretion rate feeding
 the optically thick portion of the boundary layer of the accretion
 disk is inferred to be $\dot m \approx 6.6 \times 10^{-8}$ M$_\odot$ yr$^{-1}$
 in the high state. 

 We do not have a simple explanation for this implicit
 contradiction of radio and optical/X-ray phenomenology, but we suggest
 that perhaps there are different zones in the CSM of T CrB and the radio
 only detect an outer envelope, while an inner envelope and equatorial
 density enhancement of quite higher density may be sampled in the
 optical spectra.  As a later spectral
 type, the M4-III companion of T  CrB is expected to have
 a higher wind mass loss than the M0-2 III companion of RS Oph,
 but several factors are at play, including
 metallicity \citep{Li2025}, and
\citet{Zamanov2024} found that the red giant is less luminous
 than most stars of the same spectral type and
 the R=9.70 magnitude of T CrB at quiescence with respect to that of RS Oph
 (R=11.23),   related to the distance, implies that
 the giant in RS Oph is intrinsically more luminous (mass loss
 increases with luminosity).
 \citet{Shara2018} modeled T CrB with a mass accretion rate of 2.08 
 $\times 10^{-8}$ M$_\odot$ yr $^{-1}$, which is still marginally
 compatible with both the assumption of \citet{Orlando2025},
 but also with the higher
 mass accretion rate onto the WD derived from the X-ray
 observations of recent years \citep{Luna2018, Luna2019}.  We
 note that the value
 assumed by \citet{Orlando2025} is at the low end of red giants' mass loss \citep[see][]{Schroeder2005, Li2025}.
 For comparison, RS Oph and V745 Sco the Orlando group assumed an order of magnitude
 higher density in the CSM and more than an order of magnitude greater
 density enhancement in the equatorial plane \citet{Orlando2009, Orlando2017}.
  Their model, resulting in a less strong shock in the case
 of T CrB than in the other  novae in symbiotics, rests critically
 on several parameters: 
 the assumed mass loss rate in  the red giant wind,
 its retention rate, the mass present in the equatorial enhancement,  and  
 -  {\it for the first week of the outburst} - also on the structure 
 and density of the accretion disk, which is very poorly known
 for symbiotics. An important point to notice is that accretion from
 the red giant wind is not a continuous and uniform process in
 T CrB. The system rather resembles a dwarf nova, with disk 
 instabilities and bursts of accretion \citep{Luna2019, Ilkiewicz2023}.

  Will the T CrB outburst really be less spectacular at high energy than
 RS Oph?  The X-ray and gamma-ray observations of
 the shocks in this nova (or upper limits on the gamma-rays) 
 will be a diagnostic tool to derive the parameters of the red giant companion
 and of the nova outflow, including the ejected mass. Of interest in
 the \citet{Orlando2025} model is also
 the predicted bow shock ``with a hot wake'' around the red giant, which may 
 be assessed by detecting a range of plasma temperatures.
 In short, the parameters derived for the shock in T CrB will
 be a diagnostic tool not only for the shock physics, but also
 for the physical parameters of this complex binary. 
\section{Conclusions}
Our  review of the X-ray and gamma-ray phenomenology of RS~Oph and the comparisons with other novae  known to have
 occurred in symbiotic binaries allows to draw the following 
 conclusions: 

$\bullet$ The measured (absorbed) X-ray flux of RS
 Oph peaked on days 4-5 in both the 2006 and 2021 outbursts.
  A decline followed, approximately as t$^{-1}$ until the end of the 3rd week.
 There are several uncertainties in evaluating
 the actual value of the {\it unabsorbed} flux, but
 we know that the flux in the 3-79 keV range of {\sl NuSTAR}, which
 is less sensitive to absorption, decreased in time as t$^{-1.4}$
  during the course of 26 hours, 9 days into the outburst.
 In the 4th week, the central SSS began to emerge,  making estimates of the X-ray flux due only to shocks largely uncertain, but
 the emission of shocked plasma lasted for at least 8 months in 2006
 when the nova was observable for the whole duration of the outburst.  

$\bullet$ The attempts to determine when the Sedov phase started and ended and
 to evaluate the mass of ejecta, before 2021
 have been a complicated rollercoaster, with values
 estimated  and later retracted by successive work. 
 Since there is no unique fit for the broad band X-ray data,
even with the moderate spectral resolution of {\sl NICER},
 we suggest that
 high resolution spectra should be taken at several epochs during
 the first two weeks of the outburst in order to break the degeneracy
 between  temperature, absolute flux and parameterization of the column density. 
Two papers \citep{Page2022, Orio2023} include in the
 spectral fit a rapidly cooling hot component
 and a soft component in the 0.6-1 keV range, that should
 have cooled extremely slowly both in 2006 and in 2021.  
 The peak temperature of the hottest (or only) plasma component 
 was $\simeq$25 keV soon after the outburst
 and decreased to $\simeq$2 keV by day 26.

$\bullet$ A problem that hinders 
 deriving the specific physical parameters of
 the shock is determining the time to reach  collisional
 ionization equilibrium, which depends on the electron density $n_e$, but 
 seems to differ for various atomic species. 
 It will be very valuable to obtain 
high resolution X-ray spectra already on the first day
 of the outburst of a nova occurring in a symbiotic, or as soon 
as copious X-ray flux is measured with a broad-band instrument.
 This should be followed 
 by a spectrum taken every $\simeq$ 2 days for RS Oph, or even every
 day for fast novae like T CrB or V745 Sco.
 In the first period of 1 to 4 days, the iron region close to 7\,keV in
 a luminous nova in  a symbiotic (like we hope T CrB will be)
 observed with {\sl XRISM} will show intermediate charge states approaching the He-like ion  (Fe XXIV, Fe XXIII, Fe XXII) 
allowing to measure the ionization timescale with precision.
 In the more distant future,
 {\sl NewAthena} will obtain good signal-to-noise with much shorter exposures
 than necessary with the current instruments, so that
 X-ray monitoring  with high spectral resolution
 will become much easier to obtain for a sample of novae.

$\bullet$  In RS Oph there was still evidence of shocked material  between 111
 and  250 days after the outburst, both in 1985 and in 2006.  It is
not known yet whether this was due to initial shocks and the slow cooling of
 the putative ``soft component'', or to a new shock at late epochs, like in the classical novae YZ Ret \citep{Mitrani2024} and V1716 Sco \citep{Mitrani2025},
 where shock heating must have been
ongoing, because  the high densities revealed very short cooling times. 
 
$\bullet$ Although the estimate of the {\it unabsorbed
 X-ray flux} at the beginning of the RS Oph 2021 outburst is model dependent,  
the spectral slope clearly indicates
 that the column density N(H) did not exceed 10$^{23}$ cm$^{-2}$
 while the plasma temperature reached 25 keV.
 This and the spectral slope are not compatible with 
 absorption as the reason for which the total X-ray flux was so small with
 respect to the gamma-ray flux observed with {\sl Fermi}. 
 Only a column density of N(H)$\geq 10^{24}$ cm$^{-2}$ could absorb
 the X-ray flux of the initial shock almost completely after a day; 
 on day 9 - when we have the {\sl NuSTAR} data in the hard X-ray range,    
 the column density necessary
 to block the X-ray flux would have been N(H)$\geq 10^{26}$ cm$^{-2}$,
 and again, this is not compatible with the {\it observed}
 X-ray spectrum.
 We suggest that the discrepancy in X-ray and gamma-ray
 flux  is compatible only with the hypothesis
 of at least two shocks, with the ``Fermi shock event''
 occurring very close to the red giant atmosphere. Either 
 the concomitant X-rays were totally absorbed - in practice, blocked even
 at energy of tens of keV - by the giant's dense wind
 near its atmosphere and the trail of dense
 material in the equatorial plane (so that we measured
 only the X-rays from successive shock event or events), 
 or the X-rays' emitting volume was suppressed by turbulence in the impact
 with  the thin and dense layers of much cooler plasma than
 the nova outflow, around the red giant atmosphere. 

$\bullet$ In four novae  that occurred in symbiotics having
 different orbital parameters, 
 for which we have measurements of gamma-rays in outburst,
  the {\sl Fermi} {\it peak in gamma-rays
 was always observed at the beginning of the outburst}, at the same time as the optical maximum. 
 This is not only true for the novae with the red giant within
 less than a 2 AU distance, but also for V407 Cyg, where the
 Mira companion is at a distance of about 20 AU.
 
$\bullet$  Regardless of the model, both \citet{Orio2023, Islam2024} found that the {\sl NICER} spectra, which offer better spectral
resolution than {\sl Swift},
 at least in the first two weeks are better fitted with a partially covering absorber. This seems to be due to lack of spherical symmetry in the absorbing medium: interesting, the radio observations of
 \citet{Lico2024} in the 2021 outburst 
showed a outflow in the East-West direction, with a Western lobe much brighter and more circular than the Eastern one. 

$\bullet$  The work of \citet{Orlando2025}
 modeling the expected shocks in the nearby luminous symbiotic
 nova T CrB, demonstrates 
 that the maximum temperature, peak X-ray energy, spectral slope 
 and cooling rate that will be measured are critically dependent on
 the physical parameters of the binary, including
  the red giant mass loss rate, the
 fraction of red giant wind accreted onto the WD, and the detailed structure
 of the large accretion disk. If gamma-rays are measured 
 both at TeV and GeV energy in the future T CrB outburst,
 it will be very interesting to
 verify whether the TeV gamma-ray peak is again coincident with
 the X-ray peak but not with the optical. 
 In any case, any gamma-ray detection of T CrB would indicate
 a much stronger shock than calculated by \cite{Orlando2025},
 compatible only with a
 surrounding CSM that is denser than the medium
 observable in radio observations of the system
 at quiescence, explaining the lack of forbidden lines in optical. 
  
Symbiotic novae offer a fascinating astrophysical laboratory to study
 shocks and particle acceleration. The example of RS~Oph shows that 
dense monitoring done at all wavelengths gives insight in the complex,
 yet very instructive picture of the physical phenomena in
 nova outbursts in these wide binaries. 
 A new outburst of the
 fourth Galactic recurrent nova in a symbiotic binary,
 T CrB,  which has an orbital period
 about half that of RS Oph, and is located at a third of
 the distance to RS Oph, will
 significantly constrain the models. 
  Future instruments, especially the {\sl Cherenkov Telescope ARRAY (CTA)}
 and the new X-ray facilities, for the time being {\sl XRISM} and
 in the future {\sl NewAthena}, will allow to investigate 
 these  very energetic novae with greatly
improved precision and will solve the ``mysteries''
 we outlined.

\begin{acknowledgments}
GJML is member of the CIC-CONICET (Argentina). JM acknowledges support from
the Polish National Science Center grant 2023/48/Q/ST9/00138. 
We are grateful for to the anonymous reviewer for suggestions that improved the quality of the article.
\end{acknowledgments}
%
\facilities{This research has made use of data from the NuSTAR mission, a project led by the California Institute of Technology, managed by the Jet Propulsion Laboratory, and funded by the National Aeronautics and Space Administration. Data analysis was performed using the NuSTAR Data Analysis Software (NuSTARDAS), jointly developed by the ASI Science Data Center (SSDC, Italy) and the California Institute of Technology (USA). We also used NICER data.
NICER is a 0.2-12 keV X-ray telescope operating on the International Space Station. The NICER mission and portions of the NICER science team activities are funded by NASA.}




\bibliography{rsbiblio.bib}{}
\bibliographystyle{aasjournal}



\end{document}